\documentclass[10pt,journal,compsoc]{IEEEtran}
\usepackage{amsmath}
\usepackage{graphicx}
\usepackage{xcolor}
\usepackage{subfigure}
\usepackage{diagbox}
\usepackage{multirow}
\usepackage{algorithm}
\usepackage{listings}

\usepackage{amssymb}

\DeclareMathOperator*{\argmin}{\arg\min}   

\usepackage[normalem]{ulem}
\usepackage{url}

\ifCLASSOPTIONcompsoc
  \usepackage[nocompress]{cite}
\else
  \usepackage{cite}
\fi
\ifCLASSINFOpdf
\else
\fi
\begin{document}
%
\title{MGDCF: Distance Learning via Markov Graph Diffusion for Neural Collaborative Filtering}

\author{Jun Hu, Bryan Hooi, Shengsheng Qian, Quan Fang, Changsheng Xu,~\IEEEmembership{Fellow,~IEEE}
\IEEEcompsocitemizethanks{
\IEEEcompsocthanksitem Jun Hu and Bryan Hooi are with the School of Computing, National University of Singapore, Singapore. (Corresponding author: Bryan Hooi)\protect\\
E-mail: jun.hu@nus.edu.sg; bhooi@comp.nus.edu.sg
\IEEEcompsocthanksitem 
Shengsheng Qian and Changsheng Xu are with the National Laboratory of Pattern Recognition, Institute of Automation, Chinese Academy of Sciences, Beijing, China.\protect\\
E-mail: \{shengsheng.qian,csxu\}@nlpr.ia.ac.cn
\IEEEcompsocthanksitem 
Quan Fang is with the School of Artificial Intelligence at Beijing University of Posts and Telecommunications, Beijing, China.\protect\\
E-mail: qfang@bupt.edu.cn
}
}

%
%

%

\markboth{IEEE Transactions on Knowledge and Data Engineering}{}



\IEEEtitleabstractindextext{%
\begin{abstract}

Graph Neural Networks (GNNs) have recently been utilized to build Collaborative Filtering (CF) models to predict user preferences based on historical user-item interactions.
However, there is relatively little understanding of how GNN-based CF models relate to some traditional Network Representation Learning (NRL) approaches.
In this paper, we show the equivalence between some state-of-the-art GNN-based CF models and a traditional 1-layer NRL model based on context encoding.
Based on a Markov process that trades off two types of distances, we present Markov Graph Diffusion Collaborative Filtering (MGDCF) to generalize some state-of-the-art GNN-based CF models.
Instead of considering the GNN as a trainable black box that propagates learnable user/item vertex embeddings, we treat GNNs as an untrainable Markov process that can construct constant context features of vertices for a traditional NRL model that encodes context features with a fully-connected layer.
Such simplification can help us to better understand how GNNs benefit CF models.
Especially, it helps us realize that ranking losses play crucial roles in GNN-based CF tasks.
With our proposed simple yet powerful ranking loss InfoBPR, the NRL model can still perform well without the context features constructed by GNNs.
We conduct experiments to perform detailed analysis on MGDCF.

\end{abstract}

}

\maketitle

\IEEEdisplaynontitleabstractindextext

%
\IEEEpeerreviewmaketitle


\section{Introduction}

\IEEEPARstart{P}{ersonalized} recommendation systems are widely used by Internet applications to select items that can match user preferences from a large number of candidate items.
Collaborative filtering (CF) is a technique used by a lot of personalized recommendation systems.
It assumes that if two users A and B are interested in the same item, user B is more likely to share A's preference for a different item than that of a random user.
Neural collaborative filtering (NCF) is a prevalent CF approach.
It leverages neural networks to learn vector representations of users and items based on known user-item interactions, and the similarities between vector representations of users and items can reflect the preference of users for items.
NCF has received research attention in various fields and has a number of applications in social networks~\cite{DBLP:conf/kdd/YingHCEHL18,9723516,9946432,9721542}, e-commerce~\cite{9827970,9969140,DBLP:conf/www/JiZSWWZZZL21}, micro-video platforms~\cite{DBLP:conf/mm/WeiWN0HC19,DBLP:conf/mm/WeiWN0C20}, and many other systems.

\begin{figure*}[!tbp]
\centering\includegraphics[width=6.5in]{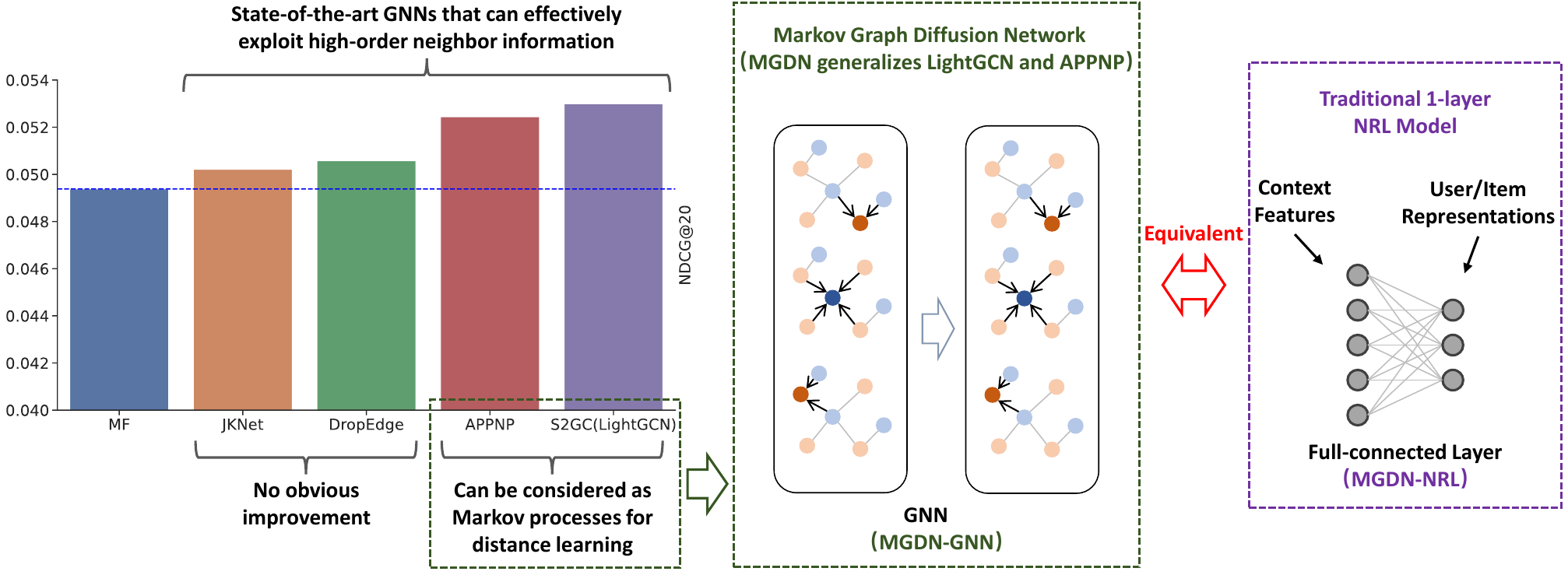}
\vspace{-2mm}
\caption{Relation between state-of-the-art GNN-based CF models and a traditional 1-layer NRL model.}
 \label{fig:intro_gnn_nrl}
\end{figure*}

NCF studies typically treat CF as a Network Representation Learning (NRL) task, which first models user-item interactions with a bipartite graph (network) and then learns low-dimensional vector representations of user and item vertices to capture implicit relations between users and items.
Traditional NRL approaches such as DeepWalk~\cite{DBLP:conf/kdd/PerozziAS14} and Matrix Factorization (MF)~\cite{DBLP:journals/computer/KorenBV09} can be used directly for the CF task.
These traditional NRL models are usually implemented as shallow neural networks.
For example, both DeepWalk and Matrix Factorization can be implemented as a 1-layer fully-connected network.
In recent years, Graph Neural Networks (GNNs) have been utilized to build NRL models, and some of them achieve promising performance on the CF task.
Compared to traditional NRL models, GNNs adopt relatively complex and deep architectures, which typically employ multiple GNN layers to iteratively aggregate information from neighbors of user/item vertices.
The left part of Fig.~\ref{fig:intro_gnn_nrl} reports the performance on the CF task of the traditional Matrix Factorization (MF)~\cite{DBLP:journals/computer/KorenBV09} approach and several 4-layer GNNs including JKNet~\cite{DBLP:conf/icml/XuLTSKJ18}, DropEdge~\cite{DBLP:conf/iclr/RongHXH20}, APPNP~\cite{DBLP:conf/iclr/KlicperaBG19}, and S$^2$GC~\cite{DBLP:conf/iclr/ZhuK21} (LightGCN~\cite{DBLP:conf/sigir/0001DWLZ020}). 
(We implement S$^2$GC as LightGCN.\footnote{ S$^2$GC assigns equal weights to neighbor information at each hop and balances the vertex's self-information with its neighbor information via a hyperparameter.
LightGCN can be considered a special case of S$^2$GC when the hyperparameter assigns equal weights to both the neighbor information at each hop and the vertex's self-information.})
While these GNNs are all powerful deep GNNs that can effectively exploit high-order neighbors, only APPNP and S$^2$GC (LightGCN) outperform MF with obvious performance improvement, suggesting that leveraging powerful GNNs will not necessarily result in superior performance on the CF task and it is thus non-trivial to design effective GNN-based CF models.

Previous research generally focuses on designing effective GNN architectures for the CF task, and there is relatively little understanding of how GNN-based CF models relate to some traditional Network Representation Learning (NRL) approaches.
In this paper, we present a unified framework named Markov Graph Diffusion Collaborative Filtering (MGDCF) to establish the \textbf{equivalence} between certain state-of-the-art GNN-based CF models and a traditional 1-layer NRL model based on context encoding.
As illustrated in the middle part of Fig.~\ref{fig:intro_gnn_nrl}, we demonstrate that the state-of-the-art GNN-based CF models APPNP and S$^2$GC (LightGCN) can be conceptualized as a Markov process that trades off distances between vertex representations.
Building on this insight, MGDCF introduces a unified model named the Markov Graph Diffusion Collaborative Network (MGDN / MGDN-GNN) to generalize (the encoders of\footnote{
MGDN, MGDN-GNN, and MGDN-NRL all refer to encoders.
Discussions on the generalization or equivalence of these models with certain GNN-based CF models focus solely on their encoders, excluding optimization.
This is because the ranking losses (see Section~\ref{sec:mf_and_gnns}) are typically applied to the encoded vertex representations, thereby allowing them to optimize different encoders interchangeably.
For simplicity, in subsequent discussions, we may omit the phrase ``the encoders of".}) these GNN-based CF models.
MGDN can be transformed into \textbf{an equivalent traditional 1-layer NRL model}, MGDN-NRL, as shown in the right part of Fig.~\ref{fig:intro_gnn_nrl}.
As with some traditional NRL approaches such as DNGR~\cite{DBLP:conf/aaai/CaoLX16}, MGDN-NRL first constructs context features for vertices without training and then encodes them into low-dimensional vertex representations via a fully-connected layer.
Such a simple architecture makes it easier to explain how GNNs benefit CF models.
In particular, employing InfoBPR, which is a powerful ranking loss of our MGDCF framework—--along with MGDN-NRL, we demonstrate that the effectiveness of GNNs may be attributed to optimization rather than regularization.
MGDN can be implemented as a heterogeneous GNN Hetero-MGDN, which is applied to a user-item graph, as well as a homogeneous GNN Homo-MGDN, which is applied to an item-item graph obtained via a sparsification technique.
We conduct experiments with the two implementations on several benchmark datasets to demonstrate the significance of our work.

In summary, the contributions of this paper are as follows:
\begin{itemize}
\item We propose Markov Graph Diffusion Collaborative Filtering (MGDCF) as a framework that establishes the equivalence between some state-of-the-art GNN-based CF models and a traditional 1-layer NRL model based on context encoding. 
MGDCF introduces a unified model named the Markov Graph Diffusion Collaborative Network (MGDN / MGDN-GNN) to generalize these GNN-based CF models, which can be further transformed into an equivalent traditional 1-layer NRL model, termed MGDN-NRL.
\item By simplifying GNN-based CF models into a 1-layer NRL model (MGDN-NRL), MGDCF makes it easier to explain how GNNs benefit CF models. 
In particular, utilizing MGDCF's powerful ranking loss, InfoBPR, we show that the impact of GNNs may come from optimization rather than regularization.
\item MGDCF can be implemented in two different forms, a heterogeneous GNN version called Hetero-MGDCF that operates on a user-item graph and a homogeneous GNN version called Homo-MGDCF for an item-item graph obtained through our sparsification technique. 
We conduct experiments with both implementations on several benchmark datasets to demonstrate the significance of our work.
\end{itemize}

\section{Related Work}

\subsection{Traditional Network Representation Learning and Graph Neural Networks}

Given a graph, Network Representation Learning (NRL) approaches aim to learn low-dimensional vector representations for vertices to capture the semantics of vertices, where the proximity of vertex representations can reflect the relationship between vertices.
Traditional NRL models typically obtain low-dimensional vertex representations by encoding graph structure information, such as the context of vertices, via shallow neural networks.
DeepWalk~\cite{DBLP:conf/kdd/PerozziAS14} utilizes random walk techniques to capture the context of vertices in the form of neighbor vertex sequences and employs a 1-layer neural network to encode the context into low-dimensional vertex representations.
Large-scale Information Network Embedding (LINE)~\cite{DBLP:conf/www/TangQWZYM15} proposes first-order and second-order proximities to model the local and global vertex context information, respectively.
LINE is also implemented as a 1-layer neural network and can be considered an efficient and scalable neural implementation of the Matrix Factorization (MF) approach.
GraRep~\cite{DBLP:conf/cikm/CaoLX15} models the context of vertices with a Positive Pointwise Mutual Information (PPMI) matrix, whose rows are high-dimensional feature vectors of vertices that explicitly capture the distribution of neighbor vertices within K hops ($K > 1$).
Using the Singular Value Decomposition (SVD)~\cite{klema1980singular} technique to factorize the PPMI matrix for linear dimension reduction, GraRep obtains low-dimensional feature vectors for vertices and shows promising performance on the NRL task.
Instead of SVD, DNGR employs an autoencoder to encode the PPMI matrix, which can be optimized in an end-to-end way.

\subsection{Graph Neural Networks}

Graph Neural Networks (GNNs) are proposed to perform deep learning on graphs.
Early GNNs such as Graph Convolutional Networks (GCNs)~\cite{DBLP:conf/iclr/KipfW17}, Graph Attention Networks (GATs)~\cite{DBLP:conf/iclr/VelickovicCCRLB18}, and GraphSage~\cite{DBLP:conf/nips/HamiltonYL17} learn vertex representations by aggregating 1-hop neighbors.
These GNNs exploit high-order neighbors via a traditional deep GNN architecture, which simply stacks multiple GNN layers with non-linear activations.
However, these models may suffer from the over-smoothing problem when more than two GNN layers are used~\cite{DBLP:conf/iclr/KipfW17,DBLP:journals/corr/abs-2008-09864,DBLP:conf/aaai/ChenLLLZS20}; therefore, they can not effectively take advantage of high-order neighbor information.
JKNet~\cite{DBLP:conf/icml/XuLTSKJ18} and DropEdge~\cite{DBLP:conf/iclr/RongHXH20} improve these models with skip connections and edge dropout, which effectively alleviate the over-smoothing problem.
Recently, various Graph Diffusion Networks (GDNs)~\cite{DBLP:conf/nips/KlicperaWG19} have been proposed to directly deal with neighbors within K hops ($K > 1$) with one GNN layer, which is more efficient.
Many powerful GNN models such as APPNP~\cite{DBLP:conf/iclr/KlicperaBG19} and S$^2$GC~\cite{DBLP:conf/iclr/ZhuK21} are GDNs.
With techniques such as Personalized PageRank~\cite{Page1999ThePC}, GDNs can obtain a propagation scheme for high-order neighbors that can avoid over-smoothing.

\subsection{Neural Collaborative Filtering}

Neural networks play an important role in modern collaborative filtering models.
SLIM~\cite{DBLP:conf/icdm/NingK11} and Multi-VAE~\cite{DBLP:conf/www/LiangKHJ18} utilize linear autoencoders or deep autoencoders to encode known interactions and reconstruct scores of unknown interactions.
Matrix factorization (MF)~\cite{DBLP:journals/computer/KorenBV09} is an essential collaborative filtering approach, which projects users and items as low-dimensional vector representations, and uses the dot product based similarities between vector representations to reflect the relevance between users and items.
MF can benefit from neural network techniques for efficient learning of vertex representations.
Neural Collaborative Filtering (NCF)~\cite{DBLP:conf/www/HeLZNHC17} suggests to replace the dot product with a similarity learned with a multilayer perceptron (MLP).
However, research ~\cite{DBLP:conf/recsys/RendleKZA20} shows that with a proper hyperparameter selection, such as a proper coefficient of L2 regularization, a simple dot product substantially outperforms the proposed learned similarities, showing that it is non-trivial to improve MF models with deep neural networks.

With the development of GNN techniques, a lot of GNN-based CF research has emerged, which can model the user-item interactions with bipartite graphs and exploit GNNs to learn representations for users and items.
GCMC~\cite{DBLP:journals/corr/BergKW17} exploits graph convolutional networks to aggregate neighbor information to build an autoencoder for matrix completion.
PinSage~\cite{DBLP:conf/kdd/YingHCEHL18} utilizes GNNs to encode both the graph structure and item features.
It designs an efficient sampling strategy based on random walks, which enables it to handle large-scale datasets.
NGCF~\cite{DBLP:conf/sigir/Wang0WFC19} is a deep GNN architecture consisting of multiple crafted graph convolutional layers, which can capture the interactions between each vertex and its neighbors.
LightGCN~\cite{DBLP:conf/sigir/0001DWLZ020} is a state-of-the-art GNN-based CF model, which removes the transformation and non-linear activations of graph convolutional layers and can effectively learn vertex representations via message propagation.
Unlike these approaches, which focus on GNN architectures, some recent research designs advanced loss functions to improve GNN-based CF models. 
Wu et al. introduce self-supervised graph learning (SGL) techniques into CF models~\cite{DBLP:conf/sigir/WuWF0CLX21} and implement them on the state-of-the-art model LightGCN.
SGL relies on graph data augmentation to generate different views for a self-supervised learning (SSL) loss, which can exploit multiple negative samples for optimization.
UltraGCN~\cite{DBLP:conf/cikm/MaoZXLWH21} replaces explicit GNN layers with a constraint loss, which can directly approximate the limit of infinite-layer graph convolutions.
UltraGCN's constraint loss also can exploit multiple negative samples.

Unlike popular GNN-based CF research such as NGCF and LightGCN, which focus on designing deep GNN architectures to exploit high-order neighbors, we investigate the GNN-based CF from the perspective of Markov processes for distance learning with a unified framework named Markov Graph Diffusion Collaborative Filtering (MGDCF).
Our unified framework can generalize state-of-the-art models such as LightGCN and APPNP, which are heterogeneous GNNs, and it can also be extended to homogeneous GNNs with our sparsification technique.
In addition, we propose InfoBPR to optimize MGDCF, which extends the widely used BPR loss to exploit multiple negative samples for better performance.

\section{Preliminary}

\subsection{Problem Definition}

We focus on featureless collaborative filtering problems, where the attributes of users and items are unknown, and only user-item interaction information is provided.
Formally, a heterogeneous user-item interaction graph $\mathcal{G}=(U \cup V, E)$ is given, where $U$ and $V$ denote the set of users and items, respectively, and $E$ is the set of edges between users and items.
Accordingly, we use $u_i$ and $v_j$ to denote the $i_{th}$ user vertex and the $j_{th}$ item vertex, respectively.
In terms of the edge set $E = \{(u_i,v_j)\}$, we use $(u_i,v_j)$ to represent the edge between $u_i$ and $v_j$, where the edges correspond to the observed interactions between users and items.
Instead of $E$, we employ a sparse adjacency matrix $\mathcal{M}^{u \rightarrow v} \in \mathbb{R}^{|U|\times |V|}$ to represent edges for computation, where each element $\mathcal{M}^{u \rightarrow v}_{ij}$ is set to 1 or 0 to denote whether the edge between the user $u_i$ and the item $v_j$ is observed. 
Given $\mathcal{G}$, our model aims to predict unobserved interactions (edges) between user vertices and item vertices, which can reflect the potential interest of users.

\subsection{Matrix Factorization and GNN-based Collaborative Filtering}\label{sec:mf_and_gnns}

Most GNN-based CF approaches such as LightGCN can be considered extensions of the Matrix Factorization (MF) approach, which is a simple yet powerful baseline approach.
Thus, we first introduce the MF approach.

\textbf{MF} aims to embed user and item vertices into a low-dimensional continuous semantic space, where the distance between user and item vertices can reflect users' interests.
Formally, MF employs an embedding table $X \in \mathbb{R}^{N \times d}$ to model the learned low-dimensional features of user and item vertices, where $N$ is the number of vertices and  $d \ll N$.

In terms of the optimization of MF, $X$ is randomly initialized and then optimized with a combined loss $\mathcal{L}_{CF}(X) = \mathcal{L}_{RANK}(X) + \Psi_{L2}\mathcal{L}_{L2}(X)$ during training.
The ranking loss $\mathcal{L}_{RANK}$, such as Binary Cross-Entropy (BCE)~\cite{DBLP:conf/www/HeLZNHC17} and Bayesian Personalized Ranking (BPR)~\cite{DBLP:conf/uai/RendleFGS09}, measures the relevance between users and items based on the dot product of learned embeddings of users and items, where a larger dot product result corresponds to higher relevance.
In addition, the L2 loss $\mathcal{L}_{L2} = ||X||_2^2$, which is multiplied with the L2 coefficient $\Psi_{L2}$, is also applied to the vertex representations $X$, which is crucial for the performance of MF~\cite{DBLP:conf/recsys/RendleKZA20}.

\textbf{GNN-based CF approaches}, which can be regarded as extensions of MF, aim to learn high-order vertex representations $Z \in \mathbb{R}^{N \times d}$ by applying a GNN model to the embedding table $X \in \mathbb{R}^{N \times d}$.
A GNN-based CF model can be represented as follows:
\begin{equation}
  Z = f_{GNN}(X, A)
\end{equation}
where $f_{GNN}$ denotes the GNN model, and $A$ is the adjacency matrix of the constructed graph.

For \textbf{optimization}, GNN-based approaches employ MF's loss function $\mathcal{L}_{CF}$ and apply it to the learned high-order vertex representations $Z$ instead of the embedding table $X$. 
That is, $\mathcal{L}_{CF}(Z) = \mathcal{L}_{RANK}(Z) + \Psi_{L2}\mathcal{L}_{L2}(Z)$.
In the official implementations of many GNN-based approaches, the L2 loss $\mathcal{L}_{L2}$ is usually applied to the GNN's parameters rather than its output $Z$.
However, we empirically find that these GNN-based models can still achieve competitive performance with $\mathcal{L}_{L2}$ applied to $Z$ instead of GNN's parameters.
To simplify the analysis, in this paper, we \textbf{only apply the L2 regularization to $Z$}.

\section{Method}

In this section, we present a unified framework named Markov Graph Diffusion Collaborative Filtering (MGDCF) to establish the equivalence between certain state-of-the-art GNN-based CF models and a traditional 1-layer NRL model based on context encoding.
Specifically, MGDCF introduces a unified model named the Markov Graph Diffusion Collaborative Network (MGDN / MGDN-GNN) to generalize these GNN-based CF models.
MGDN can be transformed into an equivalent traditional 1-layer NRL model, MGDN-NRL, which makes it easier to explain how GNNs benefit CF models.

\subsection{Generalizing State-of-the-art GNN-based CF Models with Markov Graph Diffusion Networks}\label{sec:method_generalize_mgdn}

\subsubsection{Distance Learning with Markov Graph Diffusion Networks}\label{sec:method_mgdn}

In this section, we first introduce Markov Graph Diffusion Networks (MGDNs) and then show how it is related to distance learning.

\textbf{Graph Diffusion Networks (GDNs).}
Our approach is motivated by the fact that GNNs such as APPNP and LightGCN (S$^2$GC) show superior performance on the CF task.
Different from JKNet and DropEdge, which adopt a traditional deep GNN architecture that stacks multiple GCN layers to exploit high-order neighbor information, APPNP and LightGCN are Graph Diffusion Networks (GDNs), which directly handle neighbors within K hops ($K > 1$) with one GNN layer.
Formally, a K-layer GDN can be defined by:
%
%
\begin{equation}\label{eq:gcn_coefs}
  Z_{GDN} = f_{GDN}(X, A) = (\sum_{k=0}^{K} \theta_{k} \hat{A}^{k})X
\end{equation}
where $N$ is the number of vertices, and $\hat{A} \in \mathbb{R}^{N \times N}$ is an affinity matrix of entries $\hat{A}_{ij}$ denoting the propagation weight from the $j_{th}$ vertex to the $i_{th}$ vertex.
Note that on heterogeneous graphs, the affinity matrix $\hat{A}$ defines the relevance between users and items, while on homogeneous graphs, the affinity matrix $\hat{A}$ defines the relevance between users or the relevance between items.
Many GDNs adopt GCN's normalization strategy and compute $\hat{A}$ as follows:
\begin{equation}\label{eq:common_gcn_norm}
\hat{A} = \tilde{D}^{-\frac{1}{2}}\tilde{A}\tilde{D}^{-\frac{1}{2}}  
\end{equation}
where $\tilde{A} = A + I$ and  $\tilde{D}_{ii} = \sum_{j=0}^{N} \tilde{A}_{ij}$ is the degree matrix of $\tilde{A}$.

\begin{figure*}[!t]
\centering\includegraphics[scale=0.45]{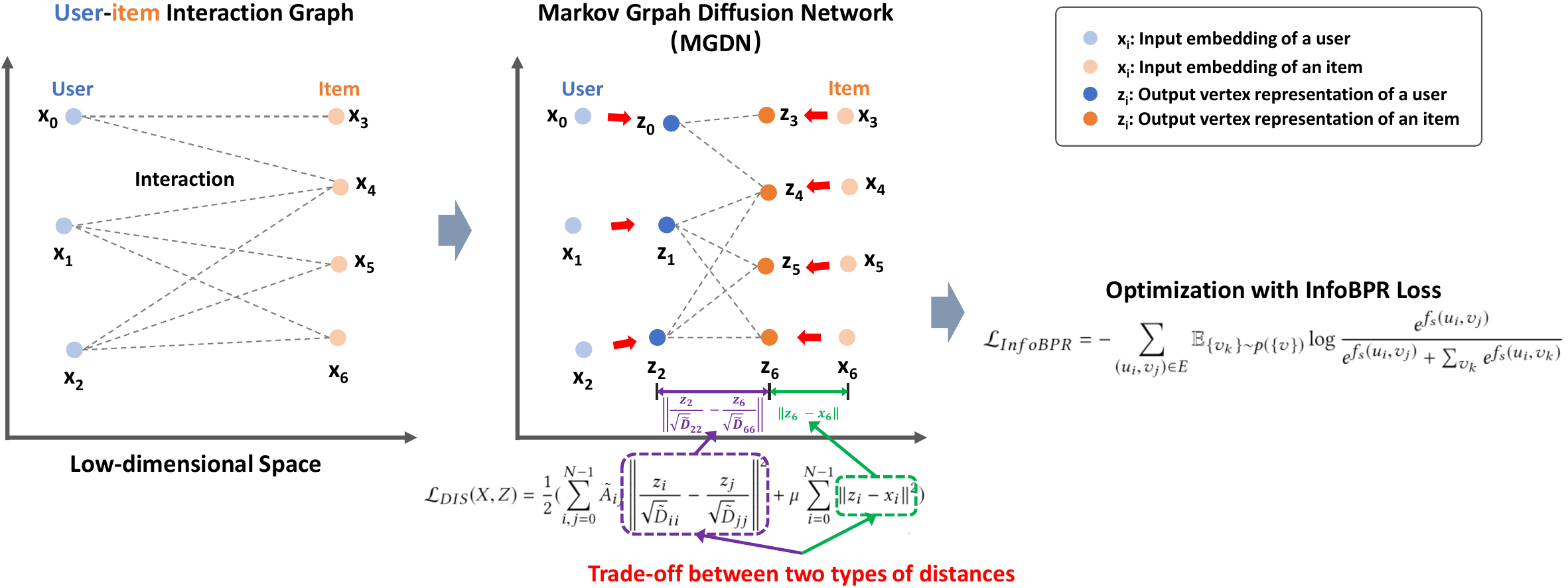}
\vspace{-3mm}
\caption{
Overall Framework of MGDCF. MGDCF uses MGDN to learn vertex representations by trading off two types of distances: (1) The left distances (in the purple dashed box) are highly related to the CF task, and minimizing it increases the smoothness of vertex representations over the graph. (2) Minimizing the right distances (in the green dashed box) is against minimizing the left distances, and it can be treated as regularization preventing over-smoothing.
}
 \label{fig:framework_hetero}
\end{figure*}

\textbf{Markov Graph Diffusion Networks (MGDNs).}
We observe that APPNP and LightGCN can be considered as a Markov Process related to distance learning of vertices, which is highly related to the CF task.
Inspired by this, we design Markov Graph Diffusion Networks (MGDNs) for distance learning by adding a constraint on the coefficients $\{\theta_{k}\}$ of Equation~\ref{eq:gcn_coefs} as follows:
\begin{equation} \label{eq:mgdn_coefs}
\begin{split}
  Z_{MGDN} 
  & = f_{MGDN}(X, A) \\
  & = (\beta^K \hat{A}^{K} + \sum_{k=0}^{K-1} \alpha \beta^{k} \hat{A}^{k}) X / \Gamma
\end{split}
\end{equation}
where $\Gamma$ is used for normalization to force $\sum_{k=0}^K \theta_{k} = 1$:
\begin{equation}
  \Gamma = \beta^K + \sum_{k=0}^{K-1} \alpha \beta^{k}
\end{equation}
Equation~\ref{eq:mgdn_coefs} can be treated as a Markov Process as follows:
\begin{equation}\label{eq:iter_init}
  H^{(0)} = X
\end{equation}
\begin{equation}\label{eq:iter_recur}
  H^{(k)} = \beta \hat{A} H^{(k-1)} + \alpha H^{(0)}
\end{equation}
\begin{equation}\label{eq:iter_final}
  Z_{MGDN} = H^{(K)} / \Gamma
\end{equation}
where Equation~\ref{eq:iter_init} and ~\ref{eq:iter_recur} correspond to the initialization and transition step of the Markov process, respectively.
The transition step (Equation~\ref{eq:iter_recur}) iteratively updates vertex representations $H^{(k)}$ through propagation. 
Both the propagation result $\hat{A} H^{(k-1)}$ and the input embeddings $H^{(0)} = X$ are incorporated into the updated vertex representations during each iteration, with the hyperparameters $\beta$ and $\alpha$ controlling their respective importance.
With the introduction of the normalization (Equation~\ref{eq:iter_final}), the Markov process would finally converges.
Specifically, when $K \to +\infty$,  we have $\lim_{K \to +\infty} \beta^K \hat{A}^{K} = 0$, and the vertex representations $Z$ converge as follows:  
\begin{equation}\label{eq:converge}
\begin{split}
  Z_{MGDN} 
  & = \lim_{K \to +\infty} \sum_{k=0}^{K-1} \alpha \beta^{k} \hat{A}^{k} X / \Gamma \\
  & = \lim_{K \to +\infty} \alpha (I - \beta^{K} \hat{A}^{K}) (I - \beta \hat{A})^{-1} X / \Gamma \\
  & = \frac{\alpha}{\Gamma} (I - \beta \hat{A})^{-1} X
\end{split}
\end{equation}

\textbf{Distance loss function.}
The converged $Z$ in Equation~\ref{eq:converge} is also the optimal solution of the Normalized Laplacian Regularization~\cite{DBLP:conf/nips/ZhouBLWS03} as follows:
\begin{equation}\label{eq:dis_loss}
\small
\mathcal{L}_{DIS}(X, Z) = \frac{1}{2} (\sum_{i,j=0}^{N-1} \tilde{A}_{ij} \left\| \frac{z_i}{\sqrt{\tilde{D}}_{ii}} - \frac{z_j}{\sqrt{\tilde{D}}_{jj}} \right\|^2 + \mu \sum_{i=0}^{N-1} \left\| z_i - x_i \right\|^2 )
\end{equation}
where the loss function $\mathcal{L}_{DIS}(X, Z)$  can be considered as a distance loss function.
As shown in Fig.~\ref{fig:framework_hetero},  $\mathcal{L}_{DIS}(X, Z)$ constrains the output vertex representations $Z$ by trading off two types of distances:
\begin{itemize}
  \item $\tilde{A}_{ij} \left\| \frac{z_i}{\sqrt{\tilde{D}}_{ii}} - \frac{z_j}{\sqrt{\tilde{D}}_{jj}} \right\|^2$ constrains the distances between the output vertex representations of linked vertices.
  Minimizing these distances increases the smoothness~\cite{DBLP:journals/corr/abs-2008-09864} of vertex representations over the graph, and they are highly related to the CF task whether on heterogeneous graphs or homogeneous graphs.
  \item $\mu \sum_{i=0}^{N-1} \left\| z_i - x_i \right\|^2$ constrains the distances between the input embedding and the output representation for each vertex.
  As shown in Fig.~\ref{fig:framework_hetero}, minimizing $\mu \sum_{i=0}^{N-1} \left\| z_i - x_i \right\|^2$ is against minimizing $\tilde{A}_{ij} \left\| \frac{z_i}{\sqrt{\tilde{D}}_{ii}} - \frac{z_j}{\sqrt{\tilde{D}}_{jj}} \right\|^2$, and it can be treated as regularization preventing over-smoothing.
\end{itemize}

\begin{figure*}[!t]
\centering\includegraphics[scale=0.43]{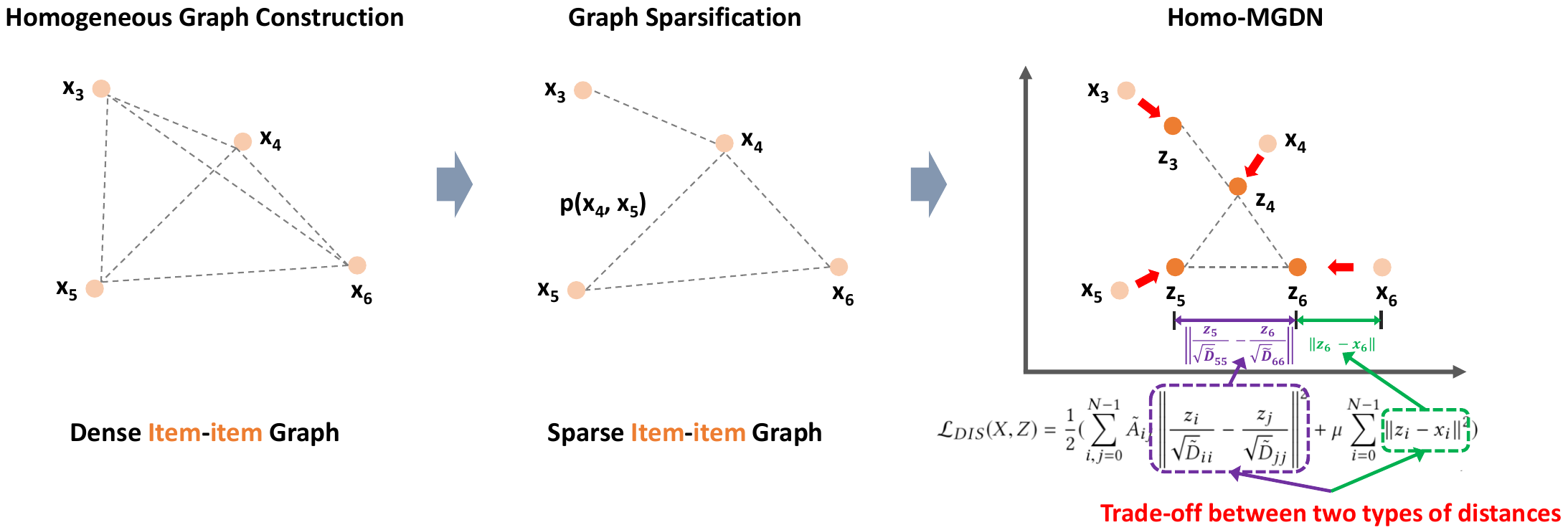}
\vspace{-5mm}
\caption{
Homogeneous MGDCF.
}
\vspace{-2mm}
 \label{fig:framework_homo}
\end{figure*}

Now, we show why the vertex representations $Z_{MGDN}$ learned by MGDN is the optimal solution of the distance loss function $\mathcal{L}_{DIS}(X, Z)$.
By converting the left part of Equation~\ref{eq:dis_loss} into the quadratic form of the normalized Laplacian matrix $\tilde{D}^{-\frac{1}{2}}(\tilde{D} - \tilde{A})\tilde{D}^{-\frac{1}{2}}$ and differentiating $\mathcal{L}_{DIS}$ with respect to $Z$, we have
\begin{equation}
\begin{split}
  \frac{\partial \mathcal{L}_{DIS}(X, Z)}{\partial Z} 
  & = \tilde{D}^{-\frac{1}{2}}(\tilde{D} - \tilde{A})\tilde{D}^{-\frac{1}{2}}Z + \mu (Z - X) \\
  & = Z - \hat{A} Z + \mu (Z - X) 
\end{split}
\end{equation}
To find $\tilde{Z}$ that can minimize $\mathcal{L}_{DIS}$ we solve the Equation below
\begin{equation}
  \tilde{Z} - \hat{A} \tilde{Z} + \mu (\tilde{Z} - X) = 0
\end{equation}
then transform it into
\begin{equation}
  \tilde{Z} - \frac{1}{1 + \mu} \hat{A} \tilde{Z} - \frac{\mu}{1 + \mu} X = 0
\end{equation}
By introducing $\beta = \frac{1}{1+\mu}$, then
\begin{equation}\label{eq:solve}
  \tilde{Z} = (1 - \beta) (I - \beta \hat{A})^{-1} X
\end{equation}
By ignoring the scaling coefficients $1 - \beta$ and $\frac{\alpha}{\Gamma}$, we find that $\tilde{Z}$, which is the optimal solution of $\mathcal{L}_{DIS}$, becomes the same as the vertex representations $Z$ learned by MGDN (Equation~\ref{eq:converge}).
Note that the reason that the scaling coefficients can be ignored is that scaling the vertex representations $Z$ will not affect the ranking of items, which is based on dot product.

\textbf{Implementations of MGDN.}
To arrive at a practical implementation, we implement MGDN in the form of a Markov Process (Equation~\ref{eq:iter_init}, \ref{eq:iter_recur}, and \ref{eq:iter_final}) since it only relies on efficient sparse matrix operations.
This allows us to implement MGDN in the message-passing framework, which is widely adopted by GNN-based collaborative filtering models. 
Especially, MGDN generalizes some state-of-the-art GNN-based collaborative filtering models such as LightGCN and APPNP:
\begin{itemize}
\item \textbf{Relation of MGDN to LightGCN} By setting $\alpha = 1.0$ and $\beta = 1.0$, Equation~\ref{eq:iter_recur} becomes $H^{(k)} = \hat{A} H^{(k-1)} + H^{(0)}$, and we have:
\begin{equation}
  Z_{LightGCN} = H^{(K)} / \Gamma = \sum_{k=0}^{K} \hat{A}^{k} X / \Gamma
\end{equation}
which corresponds to LightGCN's implementation.
Especially, such setting results in $\Gamma = K + 1$, corresponding to LightGCN's mean pooling operation.
\item \textbf{Relation of MGDN to APPNP} By treating $\alpha$ as the teleport probability of Personalized PageRank and setting $\beta = 1 - \alpha$, Equation~\ref{eq:iter_recur} becomes:
\begin{equation}
  H^{(k)} = (1 - \alpha) \hat{A} H^{(k-1)} + \alpha H^{(0)}
\end{equation}
and $\Gamma = \beta^K + (1 - \beta) \frac{\beta^{K} - 1}{\beta - 1} = 1.0$,  which corresponds to APPNP's implementation.
\end{itemize}

\subsubsection{Heterogeneous MGDN}

MGDN, when applied to heterogeneous user-item interaction graphs, is termed Hetero-MGDN.
The middle part of Fig.~\ref{fig:framework_hetero} corresponds to a Hetero-MGDN.
As shown in Fig.~\ref{fig:framework_hetero}, the distance $\tilde{A}_{ij} \left\| \frac{z_i}{\sqrt{\tilde{D}}_{ii}} - \frac{z_j}{\sqrt{\tilde{D}}_{jj}} \right\|^2$ that Hetero-MGDN tries to minimize corresponds to the distance between users and their interested items, which is directly related to the objective of CF tasks.
To implement Hetero-MGCN, we first use $A$ to denote the adjacency matrix as follows:
\begin{equation}
  A = \begin{pmatrix}
    0                                & \mathcal{M}^{u \rightarrow v}\\\
    (\mathcal{M}^{u \rightarrow v})' & 0
  \end{pmatrix}
\end{equation}
where $(\mathcal{M}^{u \rightarrow v})'$ denotes the transpose of $\mathcal{M}^{u \rightarrow v}$.
Then, we compute the affinity matrix $\hat{A}$ with Equation~\ref{eq:common_gcn_norm}.
Finally, we stack users' input embeddings and items' input embeddings to construct $X \in \mathbb{R}^{(|U| + |V|)\times d}$ and then perform MGDN with $X$ and $\hat{A}$.
The state-of-the-art approaches LightGCN and APPNP can be considered implementations of Hetero-MGDN.

\subsubsection{Homogeneous MGDN}

MGDN can also be applied to homogeneous graphs to handle CF tasks (Fig.~\ref{fig:framework_homo}).
Homogeneous MGDN (Homo-MGDN) can achieve competitive performance with Hetero-MGDN by applying MGDN to a constructed homogeneous item-item graph that captures item pairs with high co-occurrence relations.
Note that although we can also construct a homogeneous graph for users, we empirically find that constructing a homogeneous item-item graph is sufficient and we can just use the input embedding $X$ as the vertex representations $Z$ for users.
As shown in the right part of Fig.~\ref{fig:framework_homo}, the distance $\tilde{A}_{ij} \left\| \frac{z_i}{\sqrt{\tilde{D}}_{ii}} - \frac{z_j}{\sqrt{\tilde{D}}_{jj}} \right\|^2$ that Homo-MGDN tries to minimize corresponds to the distance between items that have high co-occurrence relations.

Fig.~\ref{fig:framework_homo} shows the framework of Homo-MGDN.
Based on the heterogeneous graph $\mathcal{G}$, we construct a homogeneous item-item graph $G$ to capture item pairs with high co-occurrence relations.
As with GCN, most Graph Diffusion Networks such as APPNP and S$^2$GC are based on the assumption that the edge weight between $v_i$ and $v_j$ should represent an unnormalized joint probability $p(v_i, v_j)$.
While $p(v_i, v_j)$ can be obtained (via methods such as scanning all co-occurrences of items~\cite{DBLP:conf/aaai/YaoM019}), using $p(v_i, v_j)$ might pose the potential risk of constructing a dense item-item graph.
Assuming a user, perhaps a robot, has interacted with all items, it implies that every item pair has a positive edge weight ($p(v_i, v_j) > 0)$, leading to a fully connected item-item graph, which would significantly increase the complexity of most GNNs. 
To alleviate this risk, and inspired by~\cite{DBLP:conf/nips/KlicperaWG19}, we propose a sparsification techinque, which preserves only a small fraction of edges between items with high co-occurrence relations.
(The strength of the co-occurrence relation is measured by the affinity weights, see Section~\ref{sec:sparse_proof}.)
Analogous to~\cite{DBLP:conf/nips/KlicperaWG19}, edge weight (unnormalized $p(v_i, v_j)$) is approximated to 1.0 for edges between item pairs with high co-occurrence relations and approximated to 0.0 otherwise.

Specifically, our sparsification technique first builds an item-item transition matrix $\tilde{\mathcal{M}}^{v \rightarrow v}$ to model the item-item conditional transition probability $p(v_j|v_i)$ (the left part of Fig.~\ref{fig:framework_homo}), and then approximates $p(v_i, v_j)$ with $\tilde{\mathcal{M}}^{v \rightarrow v}$ (the middle part of Fig.~\ref{fig:framework_homo}).
The sparsification involves some complex tricks and we describe the details in Section~\ref{sec:sparse_proof}.

\begin{figure*}[!tbp]
\centering\includegraphics[width=5.3in]{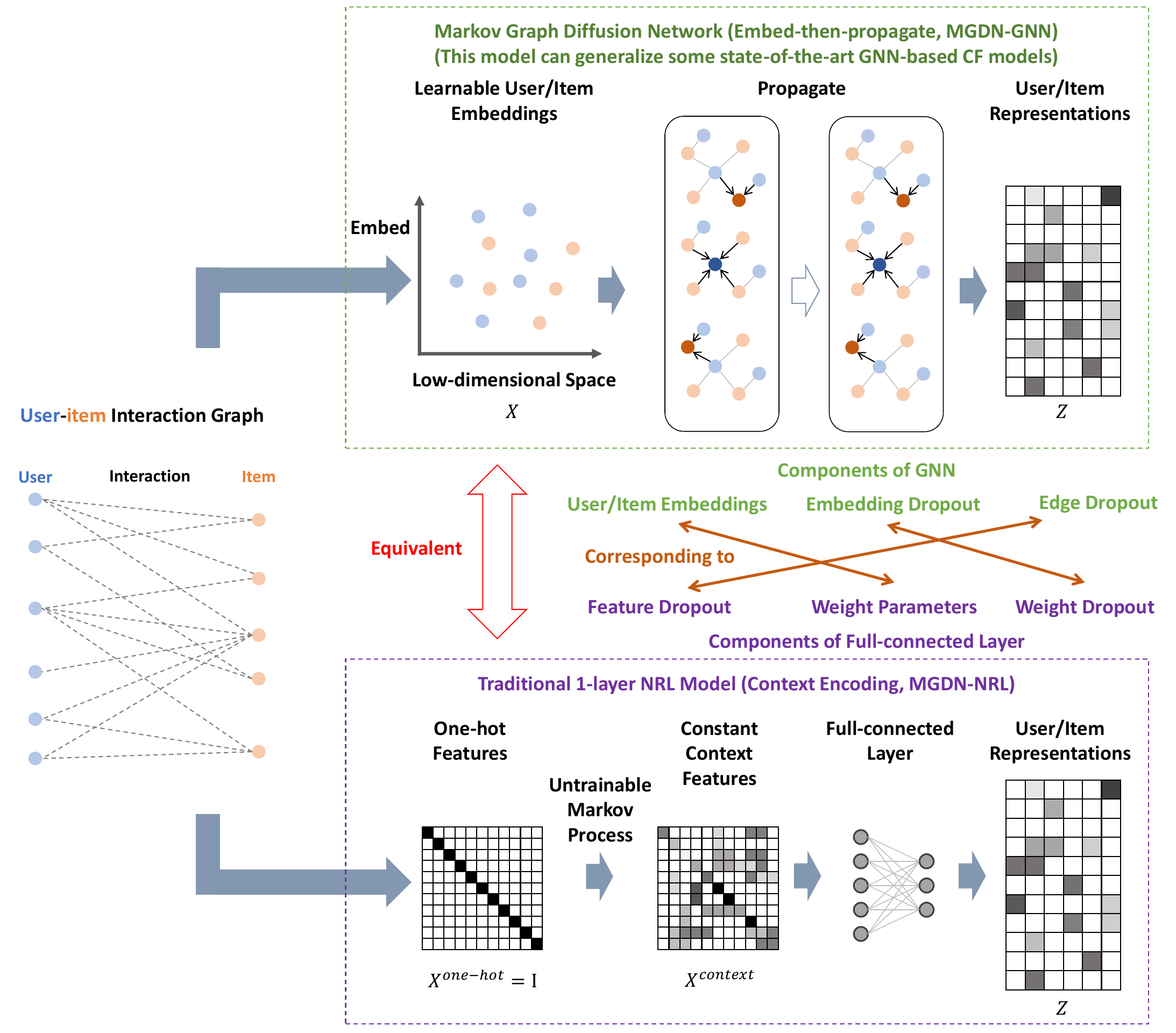}
\vspace{-2mm}
\caption{Equivalence between some state-of-the-art GNN-based CF models and a traditional 1-layer NRL model.}
\label{fig:gnn_and_nrl}
\vspace{-3mm}
 
\end{figure*}

After sparsification, we obtain the adjacency matrix $A$ for a sparsified item-item graph.
Based on $A$, we compute the affinity matrix $\hat{A}$ with Equation~\ref{eq:common_gcn_norm} and simply apply MGDN on the sparse item-item graph to learn vertex representations $Z$ for items.

\subsection{Explaining MGDN as a Traditional 1-layer NRL Model} 

In this section, we provide a simple explanation of MGDN by showing the equivalence between MGDN and a traditional 1-layer NRL model of the following form: 
\begin{equation}\label{eq:traditional_nrl_form}
  Z_{NRL} = f_{FC}(f_{CTX}(A)) = f_{FC}(X^{context})
\end{equation}
where $f_{CTX}(A) = X^{context}$ refers to some fixed function of the adjacency matrix $A$ (such as Personalized PageRank used by GraRep~\cite{DBLP:conf/cikm/CaoLX15} and DNGR~\cite{DBLP:conf/aaai/CaoLX16}) used to construct context features $X^{context}$ for vertices, and $f_{FC}$ denotes a fully-connected layer used to encode the constructed context features.
Here, the term ``context” denotes the distribution of vertices associated with the current vertex, encompassing both the vertex itself and its neighbors within $K$ hops.
(The method of constructing this association can vary according to the NRL method.)
Since MGDN can generalize state-of-the-art GNN-based CF models such as LightGCN and APPNP, our explanation also applies to these models.

We first show the limitation of the aforementioned explanation of MGDN.
Equation~\ref{eq:dis_loss} explains how MGDN refines the input user/item embeddings $X$ to obtain superior user/item vertex representations $Z$ via propagation-based distance learning.
A limitation of the explanation is that it relies on the fact that the learned user/item embeddings $X$ are already useful vertex representations for the CF task.
For example, the distance $\mu \sum_{i=0}^{N-1} \left\| z_i - x_i \right\|^2$ in Equation~\ref{eq:dis_loss} seems to be harmful to the CF task if the input user/item embeddings $X$ are terrible vertex representations.
However, as with most GNN-based CF approaches, the input user/item embeddings $X$ are randomly initialized and then optimized via back propagation.
That is, the optimization of our GNN model is treated as a black box and our explanation assumes that the optimizer can find proper input user/item embeddings $X$.
Such assumption makes the explanation incomplete.

To provide a better explanation of MGDN, we treat MGDN as a traditional 1-layer NRL model rather than an embedding-then-propagate GNN model.
Instead of encoding learnable input user/item embeddings $X$, which causes the aforementioned incomplete explanation, the traditional 1-layer NRL model encodes constant context features of vertices.
Based on Equation~\ref{eq:mgdn_coefs}, we can treat MGDN as a fully-connected layer (FC) that encodes constant context features of vertices as follows:
\begin{equation}\label{eq:gnn_and_nrl}
\begin{split}
  Z_{MGDN}
  & = f_{MGDN}(X, A) \\
  & = (\beta^K \hat{A}^{K} + \sum_{k=0}^{K-1} \alpha \beta^{k} \hat{A}^{k}) X / \Gamma \\
  & = (\beta^K \hat{A}^{K} + \sum_{k=0}^{K-1} \alpha \beta^{k} \hat{A}^{k}) I / \Gamma X\\
  & = f_{MGDN}(I, A) X\\
  & = f_{MGDN}(X^{one-hot}, A) W \\
  & = f_{FC}(X^{context}) \\
\end{split}
\end{equation}
where $X^{one-hot} = I \in \mathbb{R}^{N \times N}$ is the one-hot features of vertices, $W=X \in \mathbb{R}^{N \times d}$ is the weight of a fully-connected layer $f_{FC}$, and $X^{context}=f_{MGDN}(X^{one-hot}, A)  \in \mathbb{R}^{N \times N}$ denotes constant context features of vertices.
As shown in the lower part of Fig.~\ref{fig:gnn_and_nrl}, Equation~\ref{eq:gnn_and_nrl} treats MGDN as a traditional 1-layer NRL model mentioned by Equation~\ref{eq:traditional_nrl_form}, which consists of the following components:
\begin{itemize}
\item \textbf{Context Feature Construction:} This component is untrainable and acts on the one-hot vertex features as $X^{context}=f_{MGDN}(X^{one-hot}, A)$ to construct constant context features $X^{context}$ for vertices.
(Note that, the context is constructed via message propagation as outlined in Equation~\ref{eq:gcn_coefs}, which adopts an affinity matrix $\hat{A}$ that includes self-loops (refer to Equation~\ref{eq:common_gcn_norm}).
Therefore, each vertex is always associated with itself (due to the reachability in message propagation) and is thereby always included in its own context.)
The one-hot features $X^{one-hot}$ can be treated as the basic context features, where the context of each vertex only contains the vertex itself.
Hence, in the implicit distance loss Equation~\ref{eq:dis_loss}, the distance $\mu \sum_{i=0}^{N-1} \left\| z_i - x_i \right\|^2$ constrains $X^{context} = Z$ to preserve the basic context distribution ($X^{one-hot}$), and the distance $\tilde{A}_{ij} \left\| \frac{z_i}{\sqrt{\tilde{D}}_{ii}} - \frac{z_j}{\sqrt{\tilde{D}}_{jj}} \right\|^2$ further constrains that linked vertices $v_i$ and $v_j$ shoud have similar context features $x^{context}_i$ ($z_i$) and $x^{context}_j$ ($z_j$), respectively.
As a result, $f_{MGDN}(X^{one-hot}, A)$ can directly construct constant context features $X^{context}$ for vertices without training.
\item \textbf{Context Encoding (Fully-connected Layer):} This component is a fully-connected layer $f_{FC}$ that encodes the $N$-dimensional constant context features $X^{context}$ into $d$-dimensional vertex representations $Z$.
It is optimized by the collaborative filtering loss $\mathcal{L}_{CF}$.
\end{itemize}
Similar context encoding frameworks are adopted by traditional NRL approaches such as GraRep~\cite{DBLP:conf/cikm/CaoLX15} and DNGR~\cite{DBLP:conf/aaai/CaoLX16}, which construct context features for vertex via random surfing and encode context features using Singular Value Decomposition (SVD) and Autoencoders, respectively.
Especially, our 1-layer NRL model can be realized as a variant of DNGR, where the fully-connected layer corresponds to the encoder of DNGR's Autoencoder and the ranking loss corresponds to the decoder and the reconstruction loss of DNGR's Autoencoder.

In addition to the overall framework, some components of MGDN can be explained more simply with the equivalent components of the fully-connected layer as shown in the middle part of Fig.~\ref{fig:gnn_and_nrl}:
\begin{itemize}
\item \textbf{User/Item Embeddings (GNN) $\Leftrightarrow$ Weight Parameters (FC)}: In Equation~\ref{eq:gnn_and_nrl}, the input user/item embeddings $X$ are first moved to the outside of $f_{MGDN}$ and then used as the weight parameters $W$ of $f_{FC}$.
It becomes much easier to understand how the user/item embeddings $X$ are optimized by treating $X$ as the weight parameters $W$ of the fully-connected layer of the NRL model.
\item \textbf{Embedding Dropout (GNN) $\Leftrightarrow$ Weight Dropout (FC)}: Due to the aforementioned equivalence between embeddings (GNN) and weight parameters (FC), applying dropout on user/item embeddings is equivalent to applying dropout on the weight parameters of the fully-connected layers.
\item \textbf{Edge Dropout (GNN) $\Leftrightarrow$ Feature Dropout (FC)}: In our NRL model, the propagation is used by $f_{MGDN}(X^{one-hot}, A)$ to construct the constant context features $X^{context}$ and thus the edge dropout operation relates to the feature dropout operation of the fully-connected layer.
\end{itemize}

We use MGDN-GNN and MGDN-NRL to denote MGDN explained as an embed-then-propagate GNN framework and a traditional 1-layer NRL model, respectively.

By treating MGDN as a traditional 1-layer NRL model (MGDN-NRL) instead of an embed-then-propagate GNN framework (MGDN-GNN), the overall architecture becomes simpler, and it becomes easier to explain how GNNs benefit CF models.
Obviously, the input features $X^{context}$ and the ranking loss play crucial roles in the fully-connected layer of MGDN-NRL.
It is easy to verify the importance of the input features by comparing MGDN with MF (Section~\ref{sec:performance_analysis}), which removes the context feature construction process and directly encodes the one-hot features $X^{one-hot}$ instead of the constructed context features $X^{context}$ of vertices via a fully-connected layer.
In the next section, we explore the effect of the ranking loss of MGDN.

\subsection{Optimization with InfoBPR Loss}

To investigate the effect of the ranking loss of MGDN, we design a simple yet powerful ranking loss InfoBPR, which extends the widely used Bayesian Personalized Ranking (BPR) loss to exploit multiple negative samples.

The BPR loss~\cite{DBLP:conf/uai/RendleFGS09} is a widely used ranking loss employed by many GNN-based CF models such as NGCF and LightGCN.
BPR is defined as follows:
\begin{equation}\label{eq:bpr}
\small
  \mathcal{L}_{BPR} = - \sum_{(u_i, v_j) \in E} \mathbb{E}_{v_k \sim p(v)} \log \sigma (f_{s}(u_i, v_j) - f_{s}(u_i, v_k))
\end{equation}
where $f_{s}(u_i, v_j)$ computes the dot product between the learned vertex representations of user $u_i$ and item $v_j$ and  $\sigma(x) = \frac{1}{1 + e^{-x}}$ is the sigmoid operation.
The BPR loss can be transformed into the following form:
\begin{equation}\label{eq:bpr_other_form}
\small
\begin{split}
\mathcal{L}_{BPR} 
& = - \sum_{(u_i, v_j) \in E} \mathbb{E}_{v_k \sim p(v)} \log \frac{1}{1 + e^{-(f_s(u_i, v_j) -f_s(u_i, v_k))}}
\\
& = - \sum_{(u_i, v_j) \in E} \mathbb{E}_{v_k \sim p(v)} \log \frac{e^{f_s(u_i, v_j)}}{e^{f_s(u_i, v_j)} + e^{f_s(u_i, v_k)}}
\\
\end{split}
\end{equation}
where $v_k \sim p(v)$ samples a random item vertex from $V$ with uniform distribution.
Equation~\ref{eq:bpr_other_form} shows that the BPR loss can be considered as a softmax loss function for binary classification tasks.
Inspired by InfoNCE~\cite{DBLP:journals/corr/abs-1807-03748}, we extend the BPR loss as a multi-class classification loss, with which our model can better maximize the mutual information between users and their interested items~\cite{DBLP:conf/icml/PooleOOAT19}.
Our InfoBPR loss is defined as follows:
\begin{equation}\label{eq:info_bpr}
\small
\begin{split}
\mathcal{L}_{InfoBPR} 
& = - \hspace{-3mm}\sum_{(u_i, v_j) \in E} \hspace{-3mm} \mathbb{E}_{\{v_k\} \sim p(\{v\})} \log \frac{e^{f_s(u_i, v_j)}}{e^{f_s(u_i, v_j)} + \sum_{v_k} e^{f_s(u_i, v_k)}}
\\
\end{split}
\end{equation}
where $\mathbb{E}_{\{v_k\} \sim p(\{v\})}$ samples a set of $N_{neg}$ item vertices from $V$ with uniform distribution.
When $N_{neg} = 1$, our InfoBPR loss becomes the BPR loss. 
With InfoBPR, we optimize MGDCF with a combined loss:
\begin{equation}
  \mathcal{L}_{CF}(Z) = \mathcal{L}_{InfoBPR}(Z) + \Psi_{L2}\mathcal{L}_{L2}(Z) 
\end{equation}

\section{Theoretical Analysis}

\subsection{How MGDN Improves Collaborative Filtering}\label{sec:mgdn_improve}

In this section, we discuss how MGDN benefits collaborative filtering.
We notice that, as with Matrix Factorization (MF), MGDN does not implicitly apply regularization on the output vertex representations $Z$ to limit the output space of $Z$, which shows that the performance improvement of MGDN comes from optimization rather than regularization.
In detail, as mentioned in Section~\ref{sec:mf_and_gnns}, the same loss $\mathcal{L}_{CF}$ is applied to the vertex representations for both MF and GNN-based approaches.
Formally, we use $X^{*}$ and $Z^{*}$ to denote the vertex representations optimized by MF and MGDN, respectively, which are as follows:
\begin{equation}\label{eq:min_mf}
  X^{*} = \argmin_{X} \mathcal{L}_{CF}(X)
\end{equation}
\begin{equation}\label{eq:min_gdn}
\begin{aligned}
  Z^{*} & = \argmin_{Z} \mathcal{L}_{CF}(Z) \\
  \textrm{s.t.} \quad & Z = f_{MGDN}(X, A)    \\
\end{aligned}
\end{equation}
As with the embeddings $X$, $Z$ can be any arbitrary $d$-dimensional vector in spite of the condition $Z = f_{MGDN}(X, A)$ in Equation~\ref{eq:min_gdn}.
The reason is that $\forall Z \in \mathbb{R}^{d}$, we can find the $X$ such that $Z = f_{MGDN}(X, A)$ via the inverse function of $f_{MGDN}$ derived by Equation~\ref{eq:converge}:
\begin{equation}\label{eq:inv_func}
  X = f_{MGDN}^{-1}(Z, \hat{A}) = \frac{\Gamma}{\alpha} (I - \beta \hat{A}) Z
\end{equation}
Based on this, if we ignore the optimization problem and assume that both Equation~\ref{eq:min_mf} and Equation~\ref{eq:min_gdn} are perfectly optimized, we will have $\mathcal{L}_{CF}(X^{*}) = \mathcal{L}_{CF}(Z^{*})$ and $X^{*} = Z^{*}$, showing that MGDN and MF share the same space of possible solutions.
Compared to MF, MGDN does not change the space of possible solutions, but only changes the optimization process; therefore, the observed improvements of MGDN (Table~\ref{tab:performance_ndcg} and Table~\ref{tab:performance_recall}) can only come from the optimization rather than the regularization.

Treating MGDN as the traditional 1-layer NRL model MGDN-NRL makes it easier to understand how MGDN benefit collaborative filtering.
The MF model can be seen as a special case of MGDN-NRL (Equation~\ref{eq:gnn_and_nrl}) where $K = 0$:
\begin{equation} \label{eq:mf_nrl}
\begin{split}
  Z_{MGDN}^{K=0}
  & = f_{FC}(X^{context^{K=0}}) \\
  & = f_{FC}(f_{MGDN}^{K=0}(X^{one-hot}, A)) \\
  & = f_{FC}(X^{one-hot}) \\
  & = X^{one-hot} W \\
  & = X^{one-hot} X_{MF} \\
  & = X_{MF} \\
\end{split}
\end{equation}
where setting $K=0$ disables the MGDN as $f_{MGDN}^{K=0}(X^{one-hot}, A) = X^{one-hot}$ and the learnable user/item embeddings $X_{MF}$ of MF are used as the weight parameters of the fully-connected layer $f_{FC}$.
The special case utilizes the fully-connected layer to directly encode the one-hot features $X^{one-hot}$ rather than the context features $X^{context}$ constructed via MGDN.
That is, from the perspective of the NRL model, the only difference between MF and MGDN is the input context features of the fully-connected layer, and obviously the constructed context features $X^{context}$ contain richer information than the basic context features $X^{one-hot}$.
As a result, as shown in the experiment (Section~\ref{sec:impact_mgdn}), using $X^{context}$ (MGDN) instead of $X^{one-hot}$ (MF) results in a more efficient and effective optimization.
Note that while $X^{context}$ contains richer information than $X^{one-hot}$, it does not introduce extra information and only affects the optimization.
It is because that in features CF tasks, as with $X^{context}$, the ranking loss $\mathcal{L}_{RANK}$ is also built upon the user-item interaction graph and it is already used by the MF model.
Using $X^{context}$ allows MGDN to explicitly exploit the implicit information contained by the ranking loss $\mathcal{L}_{RANK}$, which can facilitate the optimization.
This can be further verified by the fact in the experiment (Section~\ref{sec:exp_impact_infobpr}) that MF can even achieve competitive performance with MGDN on some datasets when they are trained with the powerful ranking loss InfoBPR.

\begin{figure*}[!tp]
  \vspace{-1mm}
  \centering
  \subfigure[Gowalla]{
  \label{xxxxx} 
  \includegraphics[width=1.8in]{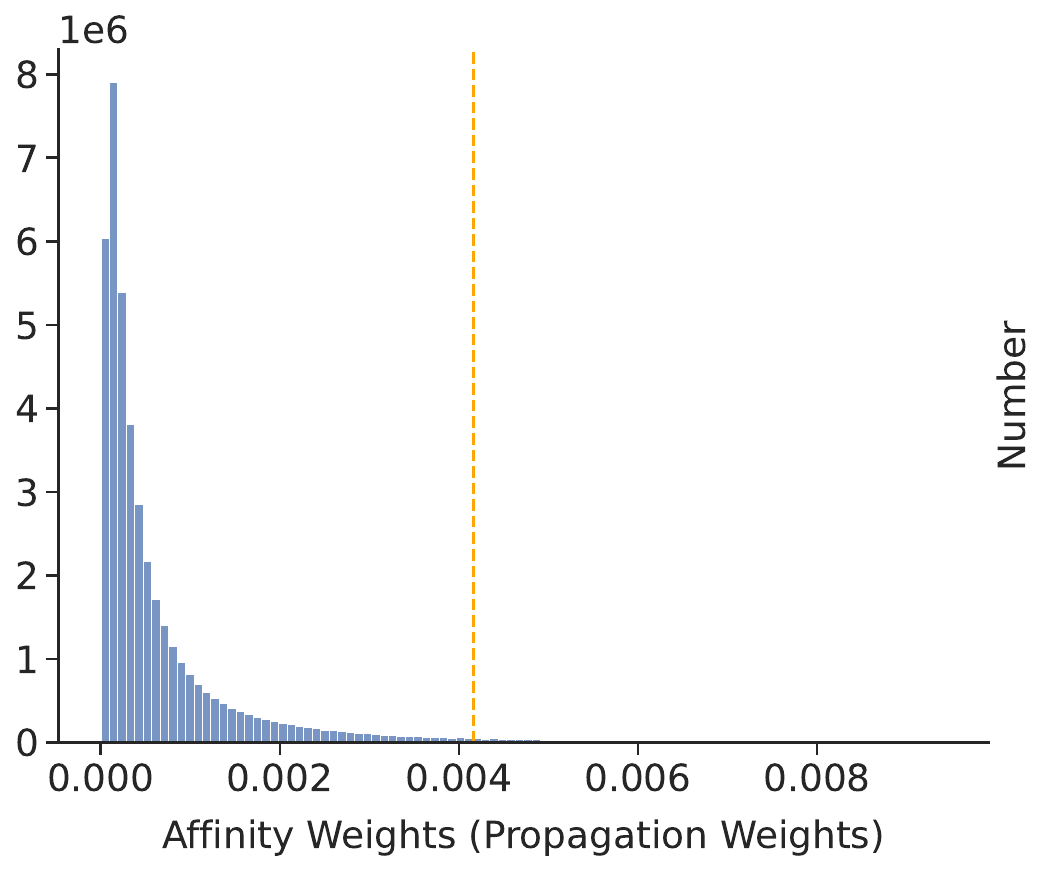}
  }
  \centering
  \subfigure[Yelp2018]{
  \label{xxxxx} 
  \includegraphics[width=1.8in]{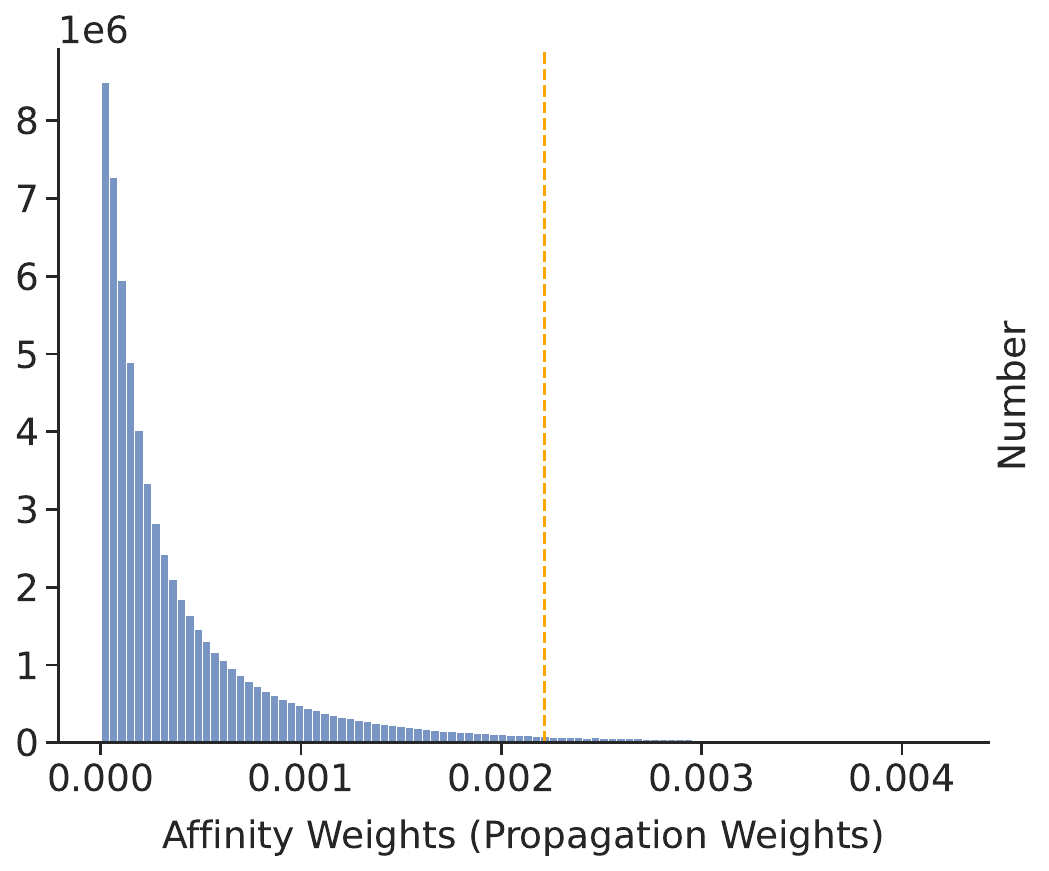}
  }  
  \centering
  \subfigure[Amazon-Book]{
  \label{xxxxx} 
  \includegraphics[width=1.8in]{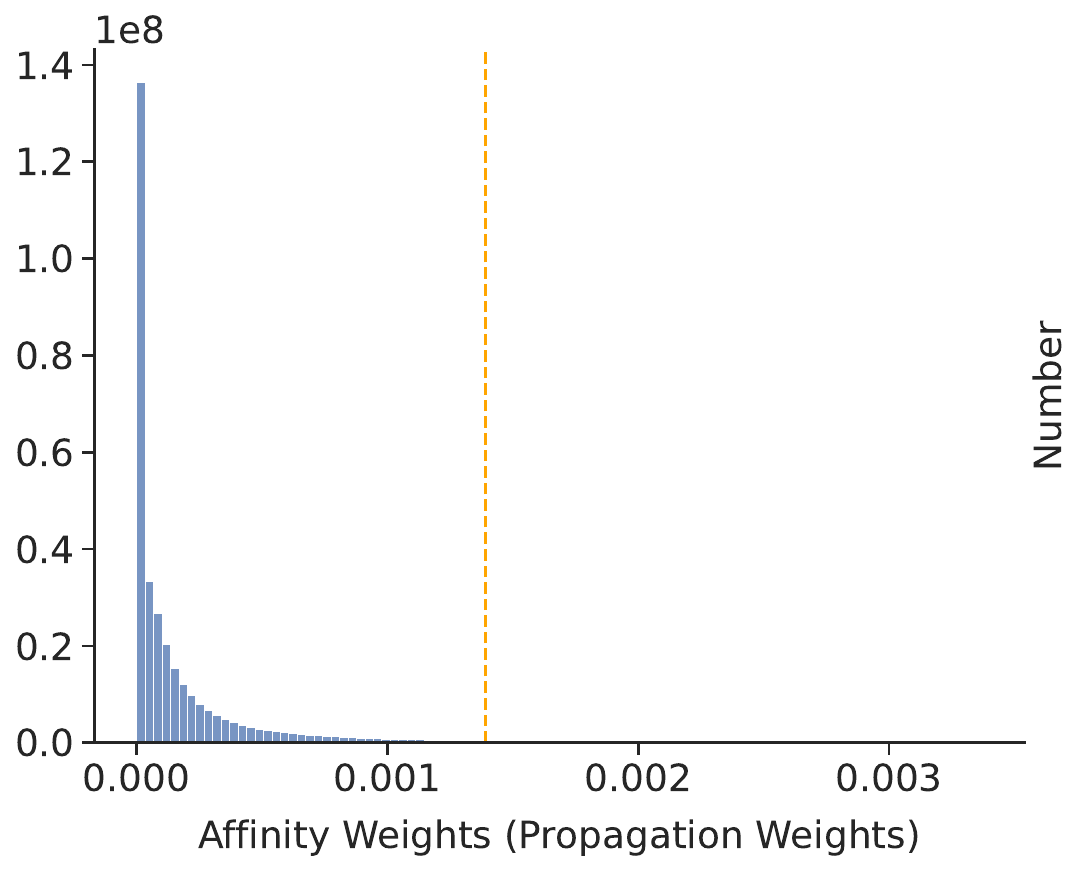}
  }  
  \vspace{-2mm}
  \caption{Distribution of affinity weights before sparsification. The orange lines denote the thresholds for sparsification.\label{fig:sparse_dist}}
  \vspace{-2mm}
\end{figure*}

\subsection{Sparsification with Transition Probability}\label{sec:sparse_proof}

In this section, we provide details about our sparsification technique.
First, we introduce how to obtain an item-item transition matrix to model the conditional transition probability between items $p(v_j|v_i)$.
Then, we introduce how it is exploited by our sparsification technique to approximate $p(v_i, v_j)$.

Given the sparse interaction matrix $\mathcal{M}^{u \rightarrow v}$, we can easily obtain the sparse user-item transition matrix as follows:
\begin{equation}\label{eq:trans_user_item}
  \tilde{\mathcal{M}}^{u \rightarrow v} = (\mathcal{D}^{u \rightarrow v})^{-1} \mathcal{M}^{u \rightarrow v}
\end{equation}
where $\mathcal{D}^{u \rightarrow v}_{ii} = \sum_{j=0}^{|V|} \mathcal{M}^{u \rightarrow v}_{ij}$ is the diagonal degree matrix of users in the heterogeneous graph.
The rows of $\mathcal{M}^{u \rightarrow v}$ are normalized and each element $\tilde{\mathcal{M}}^{u \rightarrow v}_{ij}$ of $\tilde{\mathcal{M}}^{u \rightarrow v}$ is the transition probability from user $u_i$ to item $v_j$, which is denoted as $p(v_j|u_i)$.
Similarly, we can obtain the item-user transition matrix as follows:
\begin{equation}\label{eq:trans_item_user}
  \tilde{\mathcal{M}}^{v \rightarrow u} = (\mathcal{D}^{v \rightarrow u})^{-1} \mathcal{M}^{v \rightarrow u}
\end{equation}
where $\mathcal{M}^{v \rightarrow u} = (\mathcal{M}^{u \rightarrow v})'$ is the transpose of $\mathcal{M}^{u \rightarrow v}$, and $\mathcal{D}^{v \rightarrow u}_{ii} = \sum_{j=0}^{|U|} \mathcal{M}^{v \rightarrow u}_{ij}$ is the diagonal degree matrix of items in the heterogeneous graph.

Given the two transition matrices $\tilde{\mathcal{M}}^{u \rightarrow v}$ and $\tilde{\mathcal{M}}^{v \rightarrow u}$, we can easily obtain the item-item transition matrix as follows:
\begin{equation}\label{eq:trans_item_item}
  \tilde{\mathcal{M}}^{v \rightarrow v} = \tilde{\mathcal{M}}^{v \rightarrow u} \tilde{\mathcal{M}}^{u \rightarrow v}
\end{equation}
where each element $\tilde{\mathcal{M}}^{v \rightarrow v}_{ij}$ of $\tilde{\mathcal{M}}^{v \rightarrow v}$ is a conditional probability $p(v_j|v_i)$.
It is exploited by our sparsification technique to approximate $p(v_i, v_j)$.

Our sparsification approach approximates $p(v_i, v_j)$ with the following steps:
\begin{itemize}
  \item First, we compute the affinity weights (propagation weights, reflecting the strength of the co-occurrence) between vertices based on GCN's propagation scheme, which is also used by most Graph Diffusion Networks such as APPNP and S$^2$GC.
  \item Then, we sparsify the item-item graph by keeping a small portion $(100 - s)\%$ of edges with the highest affinity weights and removing the remaining $s\%$ edges.
  \item Finally, we use 1.0 to approximate $p(v_i, v_j)$ for the kept edges.
\end{itemize}

GCN's affinity weight defined in Equation~\ref{eq:common_gcn_norm} corresponds to the following equation in the form of probability:
\begin{equation}\label{eq:gcn_prob}
  \hat{A}_{ij} = \frac{p(v_i, v_j)}{\sqrt{p(v_i)} \sqrt{p(v_j)}}
\end{equation}
GCN typically obtains the affinity weights based on $p(v_i, v_j)$, which correspond to the edge weights between items and are not provided yet.
Instead of calculating $p(v_i, v_j)$, our sparsification technique computes the affinity weights solely using $\tilde{\mathcal{M}}^{v \rightarrow v}$, whose entries represent the conditional probability $p(v_j|v_i)$.
The computation is as follows:
\begin{equation}\label{eq:hetero_to_homo}
  \hat{A} = (\tilde{\mathcal{M}}^{v \rightarrow v} \odot (\tilde{\mathcal{M}}^{v \rightarrow v})')^{\odot \frac{1}{2}}
\end{equation}
where $\tilde{\mathcal{M}}^{v \rightarrow v} \odot (\tilde{\mathcal{M}}^{v \rightarrow v})'$ denotes the Hadamard product (element-wise product) of $\tilde{\mathcal{M}}^{v \rightarrow v}$ and its transpose, and $(\cdot)^{\odot \frac{1}{2}}$ denotes the Hadamard power of an matrix.
Since $\tilde{\mathcal{M}}^{v \rightarrow v}_{ij} = p(v_j|v_i)$, the elements of $\hat{A}$ obtained by Equation~\ref{eq:hetero_to_homo} can be defined as follows:
\begin{equation}\label{eq:trans_to_gcn}
\begin{split}
\hat{A}_{ij} & = \sqrt{\tilde{\mathcal{M}}^{v \rightarrow v}_{ij} \tilde{\mathcal{M}}^{v \rightarrow v}_{ji}}\\
             & = \sqrt{p(v_j|v_i) p(v_i|v_j)} \\
             & = \sqrt{\frac{p(v_i, v_j)}{p(v_i)} \frac{p(v_i, v_j)}{p(v_j)}} \\
             & = \frac{p(v_j, v_i)}{\sqrt{p(v_i)} \sqrt{p(v_j)}}
\end{split}
\end{equation}
This shows that Equation~\ref{eq:hetero_to_homo} obtains the affinity weight of GCN's propagation scheme in Equation~\ref{eq:gcn_prob}.
As mentioned before, we then remove $s\%$ of edges based on the values of $\hat{A}_{ij}$, and use 1.0 to approximate the unnormalized $p(v_i, v_j)$ as the edge weights for the remaining edges.
After removing the self-loops, we obtain the adjacency matrix for the sparsified item-item graph, which is denoted as $A$.

Our sparsification technique can be conveniently implemented using efficient matrix operations, such as sparse matrix multiplication (Equations~\ref{eq:trans_user_item} and \ref{eq:trans_item_user}) and element-wise matrix multiplication (Equation~\ref{eq:trans_item_item}).
These operations are well-supported and optimized by standard scientific computing libraries like SciPy~\footnote{\url{https://scipy.org/}} and Numpy~\footnote{\url{https://numpy.org/}}.
Note that since sparsification is a one-time pre-processing action executed prior to training, the emphasis on its efficiency is mitigated as long as the overhead from preprocessing takes up only a small portion relative to the training process.
For example, on the Amazon dataset (see Table~\ref{tab:dataset_statistics}), our sparsification technique consumes about 1 minute, which is fast considering the entire training process requires approximately 60 minutes.

Fig.~\ref{fig:sparse_dist} shows the distribution of affinity weights before sparsification on three datasets, where the orange lines denote the thresholds for sparsification.
The affinity weights exhibit long tail distributions on both datasets, and only a small portion of edges are assigned with high affinity weights.
For the three datasets, we set $s\% = 97\%$ to remove $97\%$ of the edges, which are on the left side of the orange lines in Fig.~\ref{fig:sparse_dist}, and use 1.0 to approximate the edge weights of the remaining $3\%$ edges on the right side of the orange lines.

\section{Experiment}

This paper focuses on building a unified framework to generalize some state-of-the-art GNN-based CF models such as APPNP and LightGCN.
In this section, we conduct experiments to perform detailed analyses of each component of our unified framework.
We implement our models using tf\_geometric~\cite{DBLP:conf/mm/HuQFWZZX21}, a GNN library for TensorFlow~\footnote{\url{https://www.tensorflow.org}}. 
The source code is publicly available~\footnote{\url{https://github.com/hujunxianligong/MGDCF}}.

Note that in previous sections, we use MGDCF to refer to the overall framework, which primarily includes a GNN encoder (MGDN) and a ranking loss (InfoBPR). 
Conversely, MGDN specifically denotes the GNN encoder within the MGDCF framework, and it can be optimized with other ranking losses.
For the sake of brevity, in the experiment section, we use MGDCF and MGDN\_BPR to refer to MGDN models optimized via InfoBPR and BPR, respectively.

\begin{table}[!tp]
  \vspace{-3mm}
  \centering
  \caption{Statistics of the three datasets.}
  \vspace{-1mm}
  \scalebox{0.9}{
    \small
  \begin{tabular}{|l|c|c|c|c|}\hline
    Dataset     & Users  & Items  & Interactions & Density \\\hline
    Gowalla     & 29,858 & 40,981 & 1,027,370    & 0.084\% \\\hline
    Yelp2018    & 31,668 & 38,048 & 1,561,406    & 0.130\% \\\hline
    Amazon-Book & 52,643 & 91,599 & 2,984,108    & 0.062\% \\\hline
  \end{tabular}
  }
  \vspace{-2mm}
  \label{tab:dataset_statistics}
\end{table}

\begin{table*}[!tbp]
\centering
\caption{NDCG@20 of different approaches on three datasets.}
  \scalebox{0.62}{
    \small
\begin{tabular}{|l|c|c|c|c|c|c|c|c|c|c|c|c|}\hline
\backslashbox{Dataset}{Method}     
            & MF~\cite{DBLP:journals/computer/KorenBV09}      
                     & NGCF~\cite{DBLP:conf/sigir/Wang0WFC19}  
                              & JKNet~\cite{DBLP:conf/icml/XuLTSKJ18} 
                                       & DropEdge~\cite{DBLP:conf/iclr/RongHXH20} 
                                                  & APPNP~\cite{DBLP:conf/iclr/KlicperaBG19}   
                                                            & LightGCN~\cite{DBLP:conf/sigir/0001DWLZ020} 
                                                                     & SGL-ED~\cite{DBLP:conf/sigir/WuWF0CLX21} 
                                                                              & UltraGCN$_{Base}$~\cite{DBLP:conf/cikm/MaoZXLWH21} 
                                                                                       & UltraGCN~\cite{DBLP:conf/cikm/MaoZXLWH21}         
                                                                                                          & Hetero-MGDCF          & Homo-MGDCF      \\\hline
Gowalla     & 0.1365 & 0.1327 & 0.1391 & 0.1394   & 0.1463  & 0.1550 & --     & 0.1566 & 0.1580           & \textbf{0.1589}       & 0.1571          \\\hline
Yelp2018    & 0.0492 & 0.0461 & 0.0502 & 0.0506   & 0.0521  & 0.0530 & 0.0555 & 0.0552 & 0.0561           & 0.0572                & \textbf{0.0575} \\\hline
Amazon-Book & 0.0261 & 0.0263 & 0.0268 & 0.0270   & 0.0299  & 0.0313 & 0.0379 & 0.0393 & \textbf{0.0556}  & 0.0378                & 0.0460 \\\hline
\end{tabular}
}
\vspace{-2mm}
\label{tab:performance_ndcg}
\end{table*}

\begin{table*}[!tbp]
\centering
\caption{Recall@20 of different approaches on three datasets.}
\scalebox{0.62}{
  \small
\begin{tabular}{|l|c|c|c|c|c|c|c|c|c|c|c|c|}\hline
\backslashbox{Dataset}{Method}     
            & MF~\cite{DBLP:journals/computer/KorenBV09}      
                     & NGCF~\cite{DBLP:conf/sigir/Wang0WFC19}  
                              & JKNet~\cite{DBLP:conf/icml/XuLTSKJ18} 
                                       & DropEdge~\cite{DBLP:conf/iclr/RongHXH20} 
                                                  & APPNP~\cite{DBLP:conf/iclr/KlicperaBG19}   
                                                            & LightGCN~\cite{DBLP:conf/sigir/0001DWLZ020} 
                                                                     & SGL-ED~\cite{DBLP:conf/sigir/WuWF0CLX21} 
                                                                              & UltraGCN$_{Base}$~\cite{DBLP:conf/cikm/MaoZXLWH21} 
                                                                                       & UltraGCN~\cite{DBLP:conf/cikm/MaoZXLWH21}         
                                                                                                          & Hetero-MGDCF    & Homo-MGDCF      \\\hline
Gowalla     & 0.1591 & 0.1570 & 0.1622  & 0.1627  & 0.1708  & 0.1830 & --     &0.1845  & 0.1862           & \textbf{0.1864} & 0.1860          \\\hline
Yelp2018    & 0.0601 & 0.0566 & 0.0608  & 0.0614  & 0.0635  & 0.0649 & 0.0675 &0.0667  & 0.0683           & 0.0696          & \textbf{0.0699} \\\hline
Amazon-Book & 0.0342 & 0.0344 & 0.0343  & 0.0342  & 0.0384  & 0.0406 & 0.0478 &0.0504  & \textbf{0.0681}  & 0.0490          & 0.0566 \\\hline
\end{tabular}
}
\vspace{-1mm}
\label{tab:performance_recall}
\end{table*}

\subsection{Datasets and Baselines}
We conduct experiments with three publicly available benchmark datasets, including Gowalla, Yelp2018, and Amazon-Book~\cite{DBLP:conf/sigir/Wang0WFC19,DBLP:conf/sigir/0001DWLZ020}.
The three datasets are used by popular GNN-based CF approaches such as NGCF and LightGCN.
The statistics of the three datasets are shown in Table~\ref{tab:dataset_statistics}.

Some of the baselines are optimized with the BPR loss, including:
\begin{itemize}
  \item \textbf{MF~\cite{DBLP:journals/computer/KorenBV09}:} MF is a fundamental CF model and all the GNN-based baselines are built upon MF.
  \item \textbf{NGCF~\cite{DBLP:conf/sigir/Wang0WFC19}, JKNet~\cite{DBLP:conf/icml/XuLTSKJ18}, and DropEdge~\cite{DBLP:conf/iclr/RongHXH20}:} There GNN-based CF models adopt a traditional deep GNN architecture, which stacks multiple GNN layers together with non-linear activations and residual connections.
  \item \textbf{APPNP~\cite{DBLP:conf/iclr/KlicperaBG19} and LightGCN~\cite{DBLP:conf/sigir/0001DWLZ020}}: These models are state-of-the-art GNN-based CF models.
  They can be viewed as MGDN\_BPRs since they are MGDNs optimized with the BPR ranking loss. 
\end{itemize}
Since Hetero-MGDCF and Homo-MGDCF are optimized with InfoBPR, we also introduce baselines optimized with advanced ranking losses, including:
\begin{itemize}
  \item \textbf{SGL-ED~\cite{DBLP:conf/sigir/WuWF0CLX21}:}  SGL-ED relies on graph data augmentation to generate different views for a self-supervised learning (SSL) loss, which can exploit multiple negative samples for optimization.
  \item \textbf{UltraGCN$_{Base}$ and UltraGCN~\cite{DBLP:conf/cikm/MaoZXLWH21}:} UltraGCN directly approximates the limit of infinite-layer graph convolutions with a constraint loss, which also can take advantage of multiple negative samples.
  It maintains at least $|U|\times |V|$ negative sampling weights for all possible vertex pairs.
  As with most of our baselines and Hetero-MGDCF, UltraGCN$_{Base}$ is a heterogeneous GNN that only considers user-item relations.
  However, UltraGCN is a hybrid model that combines heterogeneous and homogeneous GNNs to incorporate different types of relations, which is not investigated by this paper.
  Therefore, the performance of UltraGCN is only reported in Table~\ref{tab:info_bpr_ndcg} and discussed in Section~\ref{sec:exp_impact_infobpr}.
  \label{sec:exp_impact_infobpr}
\end{itemize}

\subsection{Evaluation Metrics and Parameter Settings}

We employ normalized discounted cumulative gain (NDCG@20) and Recall@20 as evaluation metrics, which are also adopted by other state-of-the-art approaches such as LightGCN~\cite{DBLP:conf/sigir/0001DWLZ020} and UltraGCN~\cite{DBLP:conf/cikm/MaoZXLWH21}.

In terms of parameter settings, we aim to maintain uniformity across the official settings of most baselines and our approaches.
We primarily follow the official settings of baselines that provide settings on CF tasks, including NGCF~\cite{DBLP:conf/sigir/Wang0WFC19}, LightGCN~\cite{DBLP:conf/sigir/0001DWLZ020}, SGL-ED~\cite{DBLP:conf/sigir/WuWF0CLX21}, UltraGCN$_{Base}$~\cite{DBLP:conf/cikm/MaoZXLWH21}, and UltraGCN~\cite{DBLP:conf/cikm/MaoZXLWH21}.
Following these baselines, we set $d$ (the dimensionality of both the input embeddings $X$ and the output vertex representations $Z$) to 64 for all approaches except for NGCF, which learns $(K+1)d$-dimensional vector representations by concatenating the input embeddings and the outputs of K GNN layers.
In terms of the number of hops (which is decided by $K$), most of these baselines achieve their best performance with 3 or 4 hops of neighbors, with each achieving close performance with either 3 or 4 hops of neighbors in most cases.
Considering that Homo-MGDCF can only exploit an even number of hops of neighbors (a $K$-layer Homo-MGDCF can exploit neighbors within $K$ hops in the constructed homogeneous graphs, corresponding to neighbors within $2K$ hops in the heterogeneous graphs), we establish the number of hops as 4 to maintain uniformity across most baselines and our approaches.
Consequently, we set $K=2$ for Homo-MGDCF and $K=4$ for all other models, with the exception of SGL-ED, where we adopt $K=3$.
For SGL-ED, the official paper does not present results for $K=4$; therefore, we report the best performance indicated in the official paper, achieved at $K=3$.
For Hetero-MGDCF and Homo-MGDCF, we set $\alpha=0.1$ and $\beta=0.9$; thus, their GNN encoder MGDN is equivalent to an APPNP model, with a prior hyperparameter setting provided by the official APPNP~\cite{DBLP:conf/iclr/KlicperaBG19} paper.
For other hyperparameters such as the dropout rates, their settings may be different across different models and different datasets.
This is because different models may rely on different hyperparameter settings to achieve their best performance.
More details about the parameter settings can be found in the source code.

\subsection{Performance Analysis}\label{sec:performance_analysis}

Table~\ref{tab:performance_ndcg} and Table~\ref{tab:performance_recall} show the performance of different approaches.
From the results, we have the following observations:
\begin{itemize}
\item NGCF, JKNet, and DropEdge are deep GNNs and they perform almost the same with MF, showing that it is non-trivial to improve CF models with GNNs.
Especially, JKNet and DropEdge are powerful deep GNNs that can effectively take advantage of high-order neighbor information, and they obtain no obvious performance improvement, verifying that exploiting high-order neighbors with GNNs will not necessarily result in superior performance on the CF task.
\item APPNP and LightGCN can be considered Hetero-MGDN models and they outperform MF with significant performance improvement.
This shows that the Markov processes that trade off two types of distances can effectively help MGDN find a better optimization result. 
\item
APPNP and LightGCN are Hetero-MGDNs trained with the BPR ranking loss.
Hetero-MGDCF is also a Hetero-MGDN, but it is optimized with our InfoBPR ranking loss.
(Specifically, Hetero-MGDCF and APPNP adopt equivalent Hetero-MGDNs with the same parameter settings, and they only differ in the ranking losses).
Hetero-MGDCF demonstrates substantial performance improvement over APPNP and LightGCN, highlighting the superiority of our InfoBPR loss over the BPR loss.
Moreover, this emphasizes the crucial role of ranking losses in GNN-based CF tasks.
\item 
SGL-ED and UltraGCN$_{Base}$ are heterogeneous GNN models and they outperform other heterogeneous baselines with dramatic improvement via complex ranking losses.
This reinforces the significance of ranking losses in the optimization of CF models.
Both SGL-ED and UltraGCN$_{Base}$ employ complex ranking losses.
SGL-ED's ranking loss relies on graph data augmentation to generate different views for self-supervised learning (SSL) and utilizes multiple negative samples for optimization.
UltraGCN$_{Base}$'s ranking loss carefully assigns different weights to different negative samples, which maintains $|U|\times |V|$ negative sampling weights for all possible vertex pairs.
In contrast to these complex ranking losses, our InfoBPR loss simply extends the BPR loss. 
Despite its simplicity, with InfoBPR, our heterogeneous GNN Hetero-MGDCF achieves competitive performance against SGL-ED and UltraGCN$_{Base}$ on the three datasets, showing that our InfoBPR loss is simple yet powerful.
\item Homo-MGDCF applies our GNN model MGDN to homogeneous graphs and achieves competitive or superior performance compared to Hetero-MGDCF across different datasets, showing that MGDCF can also be effectively applied to homogeneous graphs.
Especially, Homo-MGDCF outperforms Hetero-MGDCF with significant improvement on the Amazon-Book dataset, showing that our sparsification approach used for homogeneous graph construction may implicitly suppress some noisy relational information in graphs.
\item
UltraGCN is a hybrid model that combines heterogeneous and homogeneous GNNs to incorporate different types of relations, which is not investigated by this paper.
This hybrid approach results in performance improvements over its heterogeneous counterpart, UltraGCN$_{Base}$.
Nonetheless, our non-hybrid models, either Hetero-MGDCF or Homo-MGDCF, can still outperform or match UltraGCN's performance on two out of three datasets.
In this paper, we focus solely on non-hybrid GNNs, whether heterogeneous or homogeneous, and we plan to consider hybrid models that combine heterogeneous and homogeneous GNNs in future work.

\end{itemize}

\begin{figure}[!tbp]
  \vspace{-4mm}
  \centering
  \subfigure[Gowalla]{
  \label{xxxxx} 
  \includegraphics[width=2.8in]{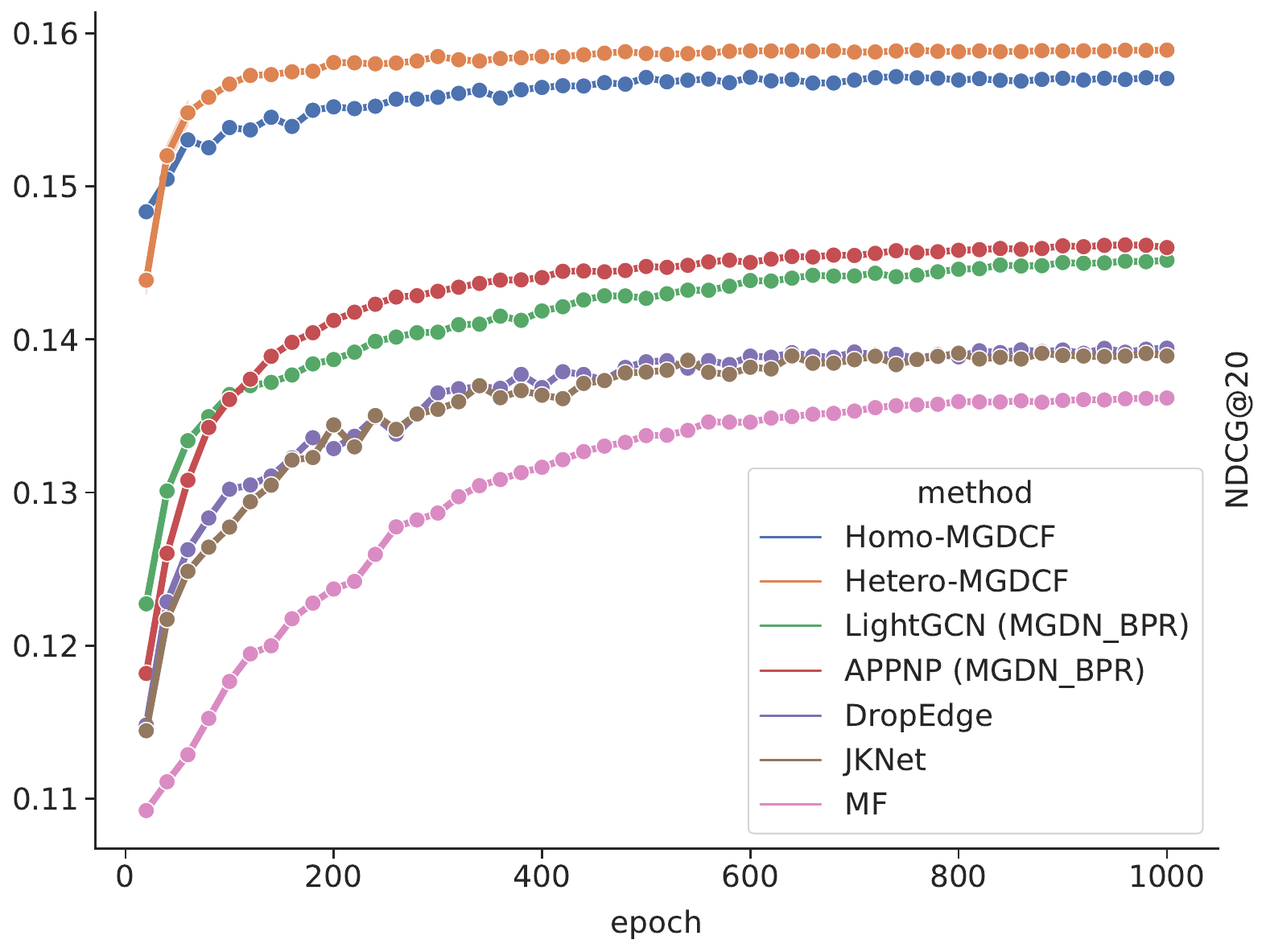}
  }  
  \centering
  \subfigure[Yelp2018]{
  \label{xxxxx} 
  \includegraphics[width=2.8in]{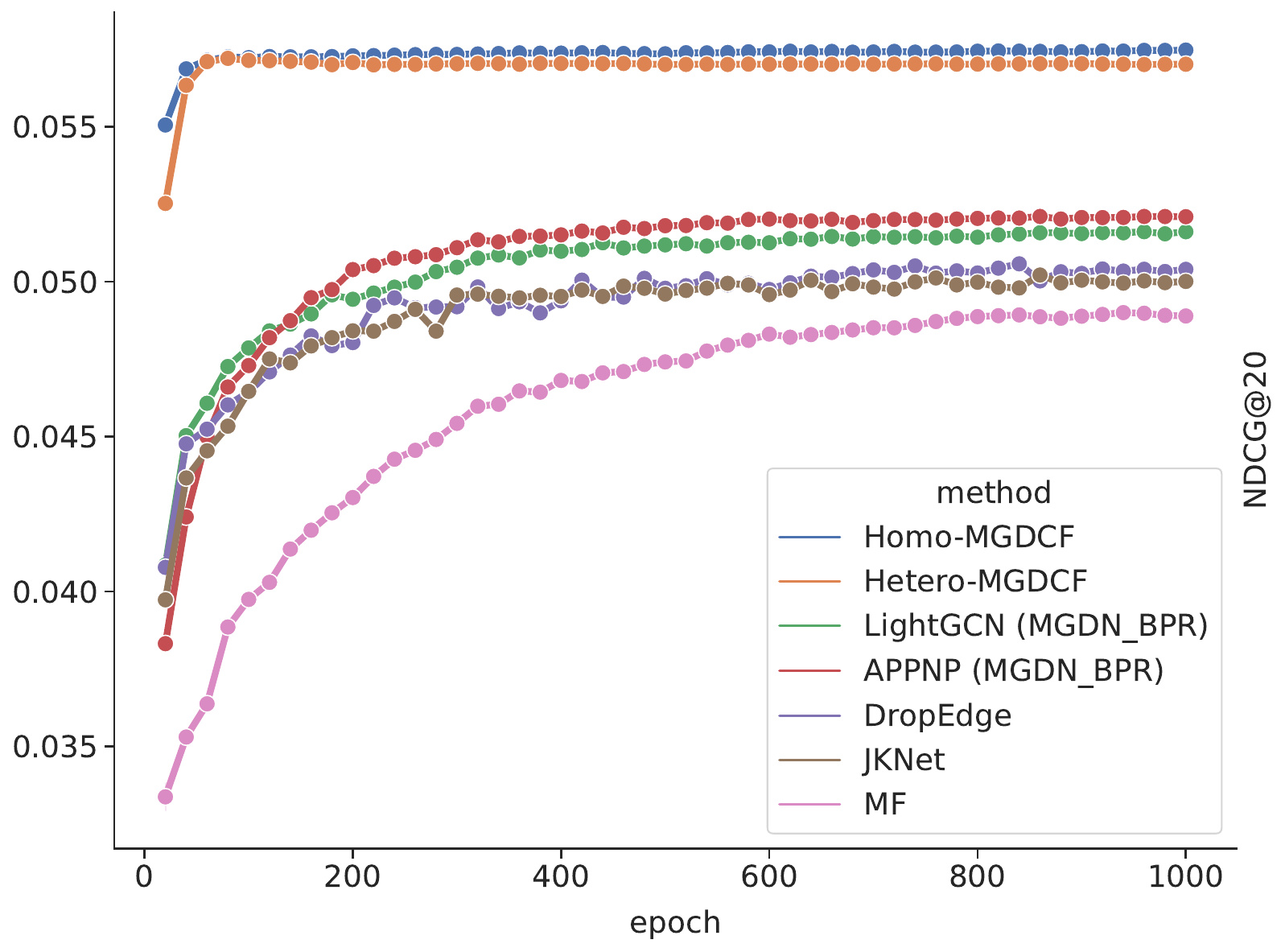}
  }
  \centering
  \subfigure[Amazon-Book]{
  \label{xxxxx} 
  \includegraphics[width=2.8in]{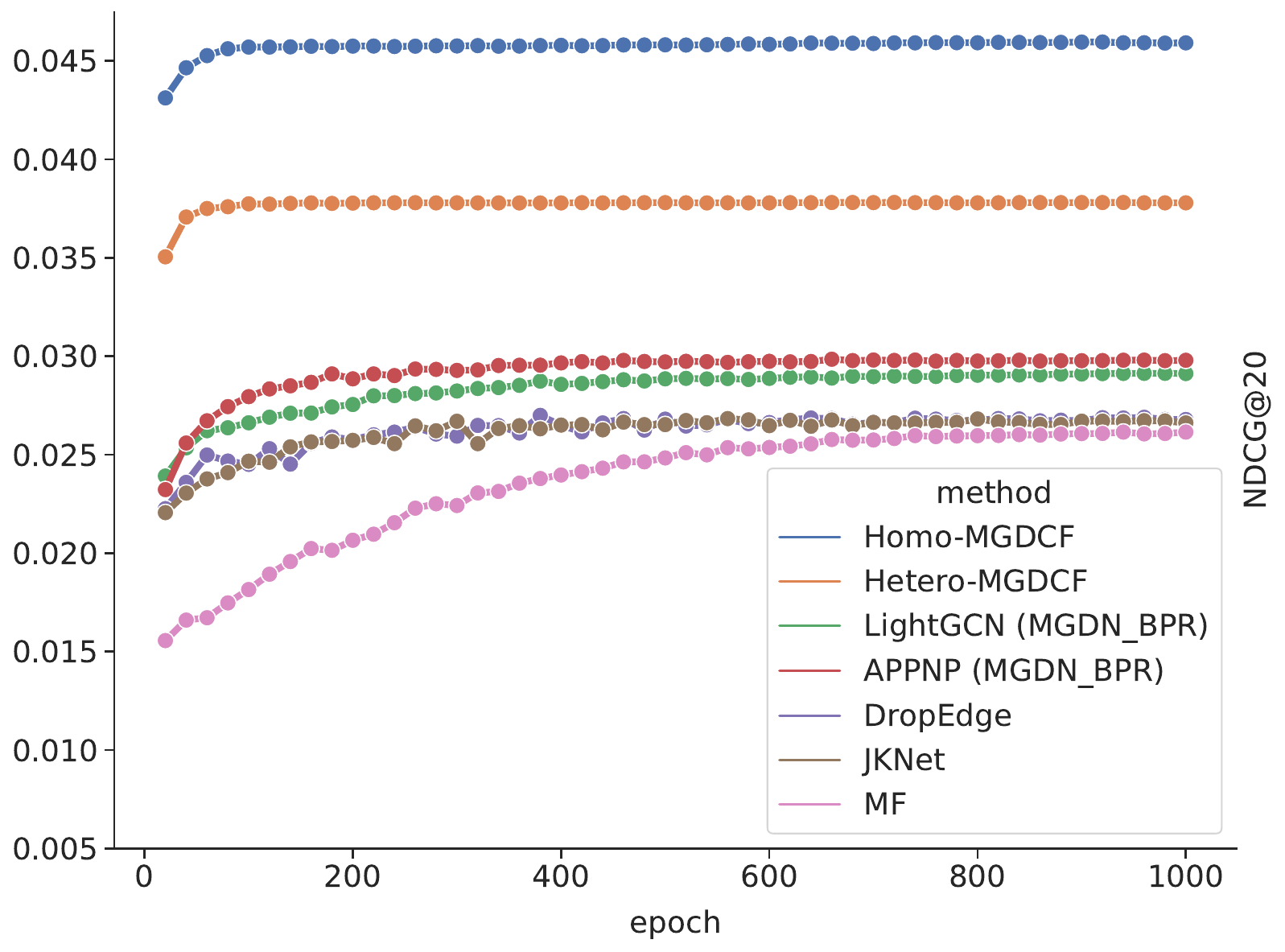}
  }
  \vspace{-2mm}
  \caption{Performance (NDCG@20) after different epochs of training.\label{fig:mf_gnn_speed}}
  \vspace{-4mm}
\end{figure}

\begin{table*}[!tbp]
  \centering
  \caption{Impact of InfoBPR Loss (NDGC@20).}
  \scalebox{0.8}{
    \small
  \begin{tabular}{|l|c|c|c|c|c|c|c|c|c|}\hline
    \backslashbox{Dataset}{Method}  
              & MF~\cite{DBLP:journals/computer/KorenBV09}
                       & MF-MultiBPR    & SGL-ED~\cite{DBLP:conf/sigir/WuWF0CLX21}         
                                                         & UltraGCN$_{Base}$~\cite{DBLP:conf/cikm/MaoZXLWH21}
                                                                                       & UltraGCN~\cite{DBLP:conf/cikm/MaoZXLWH21}         
                                                                                                          & MF-InfoBPR      & Hetero-MGDCF    & Homo-MGDCF      \\\hline
  Loss        & BPR    & MultiBPR       & SSL + BPR      & \multicolumn{2}{c|}{Constraint + BCE}          & \multicolumn{3}{c|}{InfoBPR}                   \\\hline  
  K           & \multicolumn{2}{c|}{--} & 3              & \multicolumn{2}{c|}{--}                        & --              & 3               & 2               \\\hline
  Gowalla     & 0.1365 & 0.1360         &   --           & 0.1566                      & 0.1580           & 0.1547          & \textbf{0.1595} & 0.1571          \\\hline
  Yelp2018    & 0.0492 & 0.0515         & 0.0555         & 0.0552                      & 0.0561           & \textbf{0.0581} & 0.0574          & 0.0575 \\\hline
  Amazon-Book & 0.0261 & 0.0284         & 0.0379         & 0.0393                      & \textbf{0.0556}  & 0.0373          & 0.0371          & 0.0460           \\\hline
  \end{tabular}
  }
  \label{tab:info_bpr_ndcg}
  \end{table*}

\vspace{-2mm}
\subsection{Detailed Analysis}

\subsubsection{Impact of MGDN}\label{sec:impact_mgdn}

We show the performance of MF and several GNN-based CF models during optimization in Fig.~\ref{fig:mf_gnn_speed}.
In this section, we focus on approaches optimized with BPR loss, including MF, JKNet, DropEdge, APPNP (MGDN\_BPR), and LightGCN (MGDN\_BPR).
We implement APPNP and LightGCN as Hetero-MGDN models and adopt the loss function described in Section~\ref{sec:mf_and_gnns}, which is different from the official implementation and results in slight differences in performance.

In Fig.~\ref{fig:mf_gnn_speed}, all the GNN-based models converge faster than MF, showing that GNNs may implicitly affect the optimization.
However, the final performance of JKNet and DropEdge is only slightly higher than MF, showing that the performance of these GNN models is dominated by the loss function $\mathcal{L}_{CF}$.
APPNP and LightGCN are Hetero-MGDN models, and they lead to not only a faster optimization, but also a better optimization result.
This shows that although MGDN does not apply any regularization on the vertex representation $Z$ (Section~\ref{sec:mgdn_improve}), the distance loss $\mathcal{L}_{DIS}$ (Equation~\ref{eq:dis_loss}) may implicitly affect the optimization and result in a more efficient and effective optimization.

\subsubsection{Impact of InfoBPR loss}\label{sec:exp_impact_infobpr}

As shown in Fig.~\ref{fig:mf_gnn_speed}, Hetero-MGDCF and Homo-MGDCF, which are optimized with InfoBPR loss, converge extremely faster than other approaches, which are optimized with the BPR loss.
(Specifically, both Hetero-MGDCF and APPNP employ equivalent Hetero-MGDNs, adhering to the same parameter settings, with the only difference residing in the ranking losses they employ).
Moreover, Fig.~\ref{fig:mf_gnn_speed} shows that InfoBPR also enables Hetero-MGDCF and Homo-MGDCF to obtain superior optimization results with significant performance improvement.
This shows that InfoBPR can result in a more efficient and effective optimization than BPR.

\begin{figure*}[!tbp]
  \vspace{-1mm}
  \centering
  \subfigure[Gowalla]{
  \label{xxxxx} 
  \includegraphics[width=2.2in]{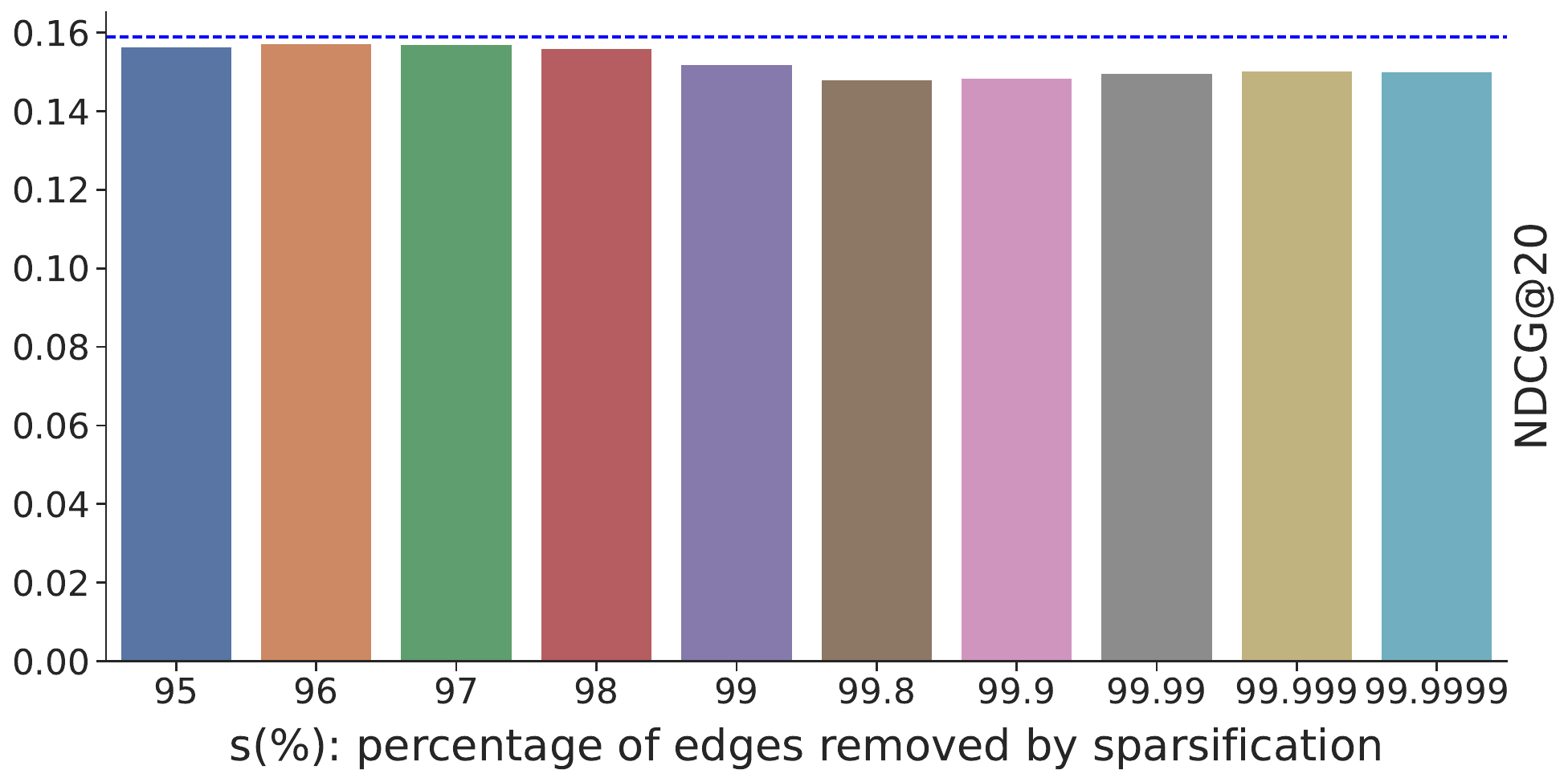}
  }  
  \centering
  \subfigure[Yelp2018]{
  \label{xxxxx} 
  \includegraphics[width=2.2in]{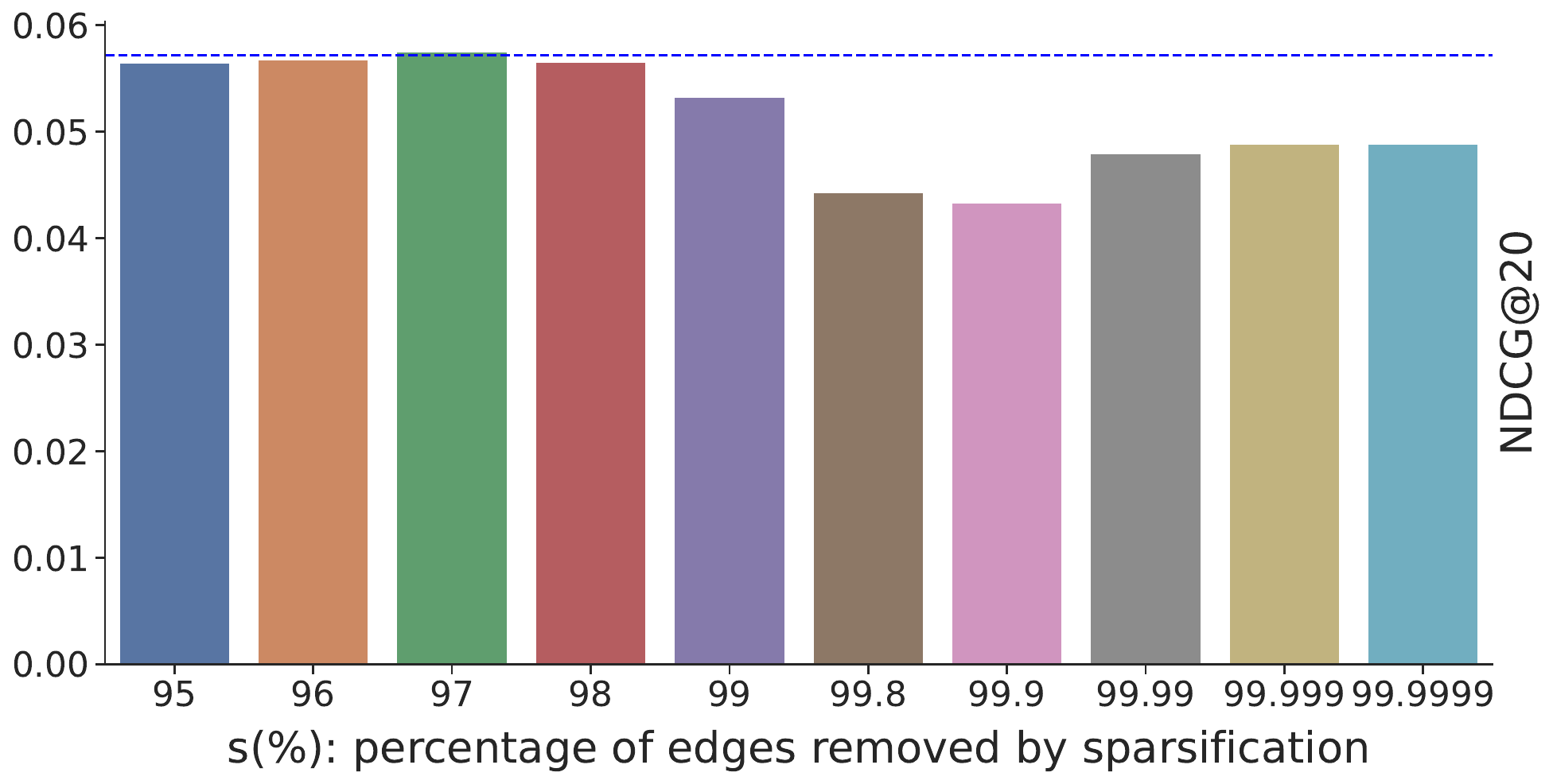}
  }
  \centering
  \subfigure[Amazon-Book]{
  \label{xxxxx} 
  \includegraphics[width=2.2in]{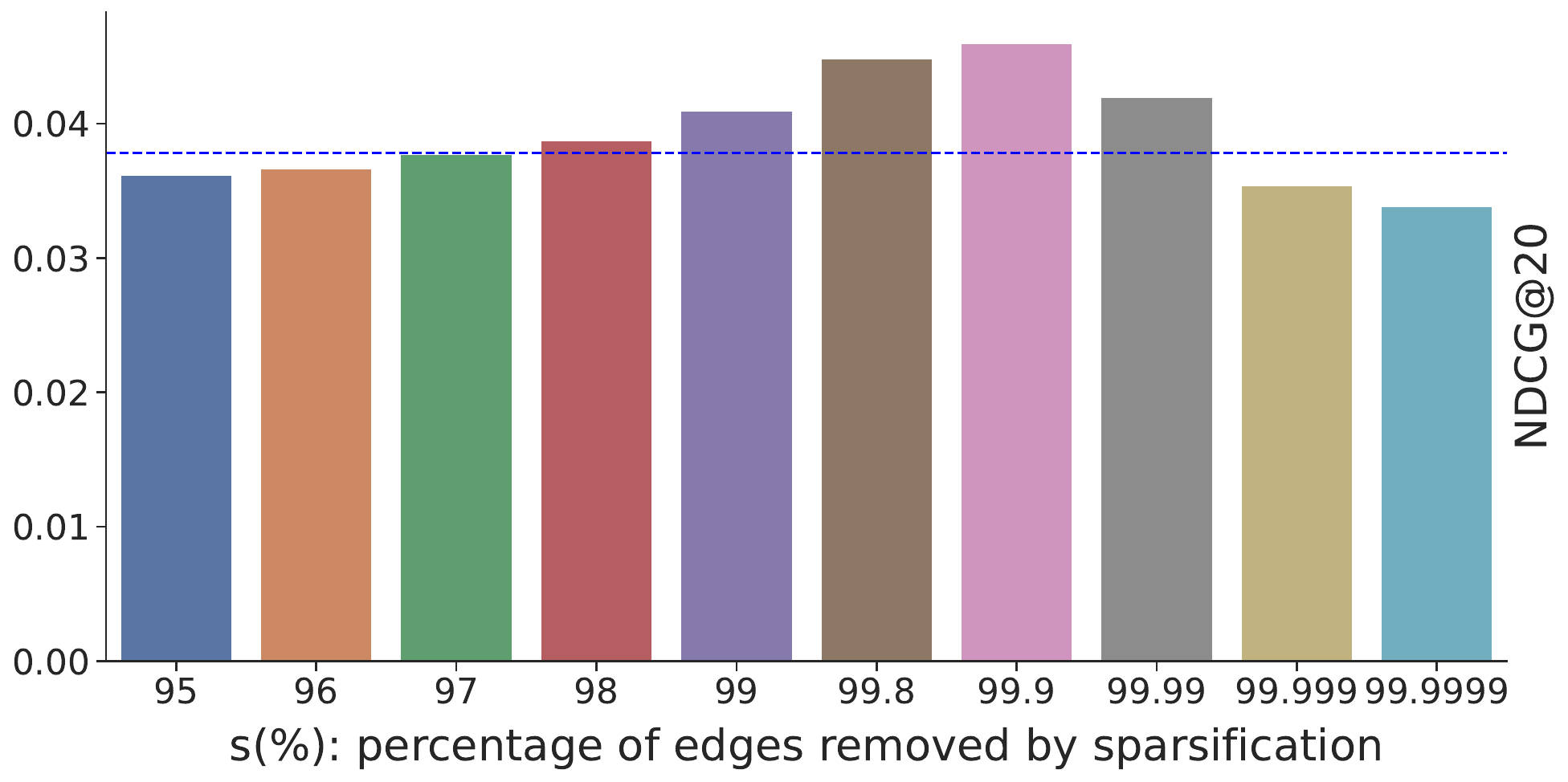}
  }
  \vspace{-2mm}
  \caption{Impact of homogeneous graph construction for Homo-MGDCF (NDCG@20). The blue lines represent the performance of Hetero-MGDCF.
  \label{fig:sparsification}}
\end{figure*}

We also compare InfoBPR with other loss functions besides BPR.
Table~\ref{tab:info_bpr_ndcg} depicts the performance of models optimized with advanced loss functions supporting multiple negative samples, including MF-MultiBPR, MF-InfoBPR, SGL-ED, UltraGCN$_{Base}$, UltraGCN, Hetero-MGDCF, and Homo-MGDCF.
We introduce a naive variant of the BPR loss, termed MultiBPR, which simply samples multiple negatives to form multiple triplets and then applies the BPR loss to each triplet.
MF-MultiBPR and MF-InfoBPR refer to the MF model optimized with MultiBPR and InfoBPR loss, respectively.
SGL-ED, UltraGCN$_{Base}$, and UltraGCN adopt complex ranking losses: SGL-ED introduces a self-supervised learning (SSL) loss, which relies on extra data augmentations and needs to be used together with the BPR loss; UltraGCN$_{Base}$ and UltraGCN adopt a complex ranking loss that carefully assigns different weights for different negative samples.
In contrast, our InfoBPR loss is much simpler and it simply extends the BPR loss to support multiple negative samples.
In Table~\ref{tab:info_bpr_ndcg}, we report performance under $K = 3$  due to the lack of official results for SGL-ED under $K=4$.
From the results, we observe:
\begin{itemize}
\item 
MF-MultiBPR can only match or slightly outperform MF.
In contrast, MF-InfoBPR shows significant performance improvement over both MF and MF-MultiBPR.
Especially, the performance improvement of MF-InfoBPR over MF is about 5.5 times the  performance improvement of MF-MultiBPR over MF.
It shows that it is non-trivial to extend BPR to support multiple negative samples, and shows the superiority of InfoBPR over the native solution MultiBPR.
\item Despite InfoBPR's simplicity, our models Hetero-MGDCF and Homo-MGDCF can achieve competitive performance with UltraGCN$_{Base}$ and SGL-ED across different datasets, and MF-InfoBPR can even beat all other methods on the Yelp2018 dataset, showing that InfoBPR is a simple yet powerful ranking loss.
In terms of UltraGCN, it is a hybrid model that combines heterogeneous and homogeneous GNNs to incorporate different types of relations, which is not investigated by this paper.
Although such a combination brings UltraGCN superior performance on the Amazon-Book dataset, our Hetero-MGDCF and Homo-MGDCF still can match UltraGCN on Gowalla and Yelp2018.
In this paper, we only investigate heterogeneous or homogeneous GNNs, and we will consider hybrid models that combine heterogeneous and homogeneous GNNs in the future.
\item 
MF-InfoBPR can achieve competitive performance with Hetero-MGDCF (i.e., Hetero-MGDN optimized via InfoBPR) on the Amazon-Book dataset and even outperform Hetero-MGDCF on the Yelp2018 dataset.
This is different when using the less powerful BPR loss, where Hetero-MGDN optimized via BPR (i.e., APPNP/LightGCN in Table~\ref{tab:performance_ndcg}) outperforms MF with dramatic performance improvement.
This observation verifies our opiniton mentioned in Section~\ref{sec:mgdn_improve} that the performance improvement of MGDN comes from optimization rather than regularization.
\end{itemize}

\begin{figure*}[!tp]
\centering
\subfigure[Impact of $\alpha$\label{fig:impact_param_alpha}]{
\includegraphics[width=1.8in]{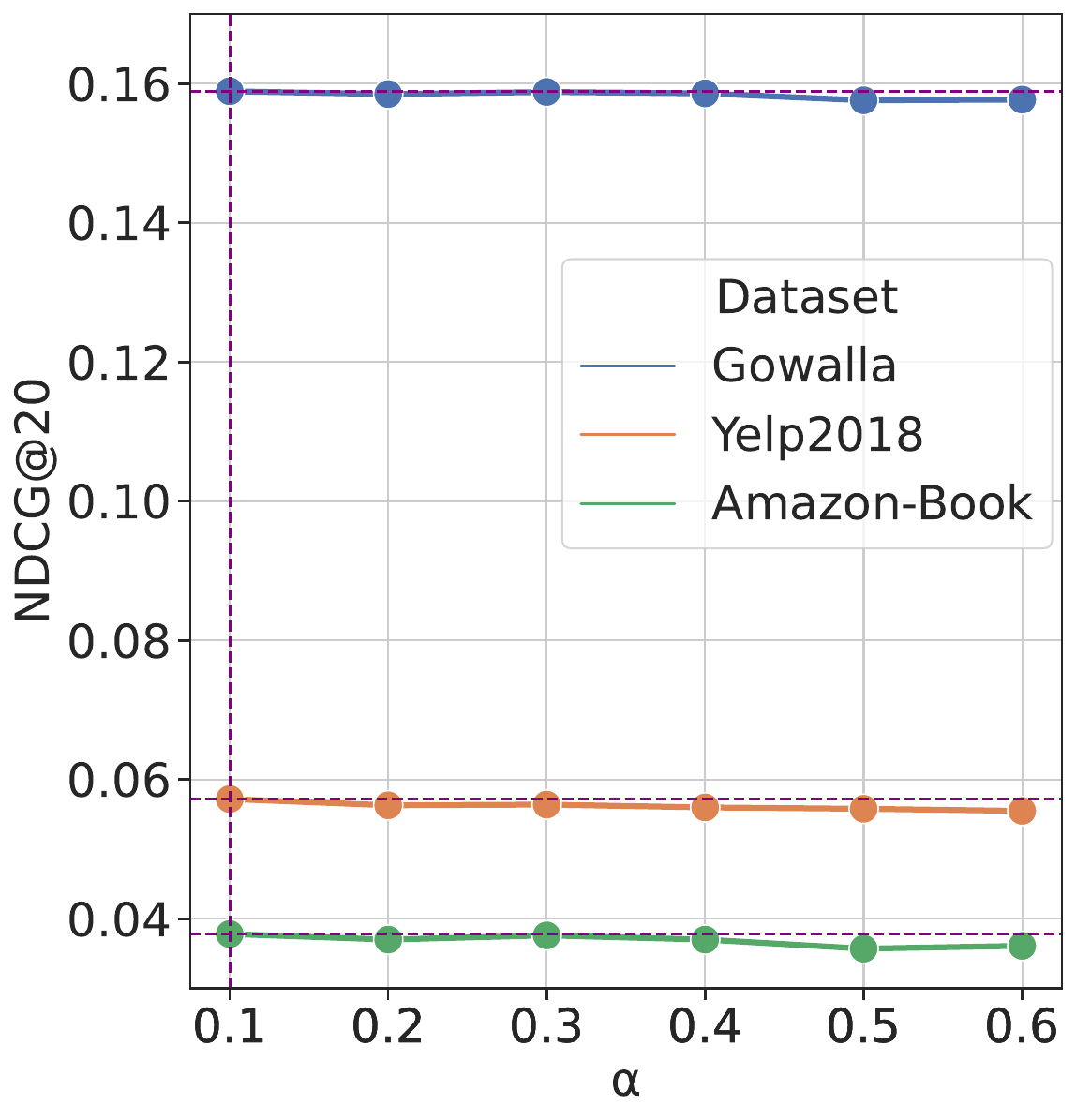}
}
\hspace{5mm}
\centering
\subfigure[Impact of $\beta$\label{fig:impact_param_beta}]{
\includegraphics[width=1.8in]{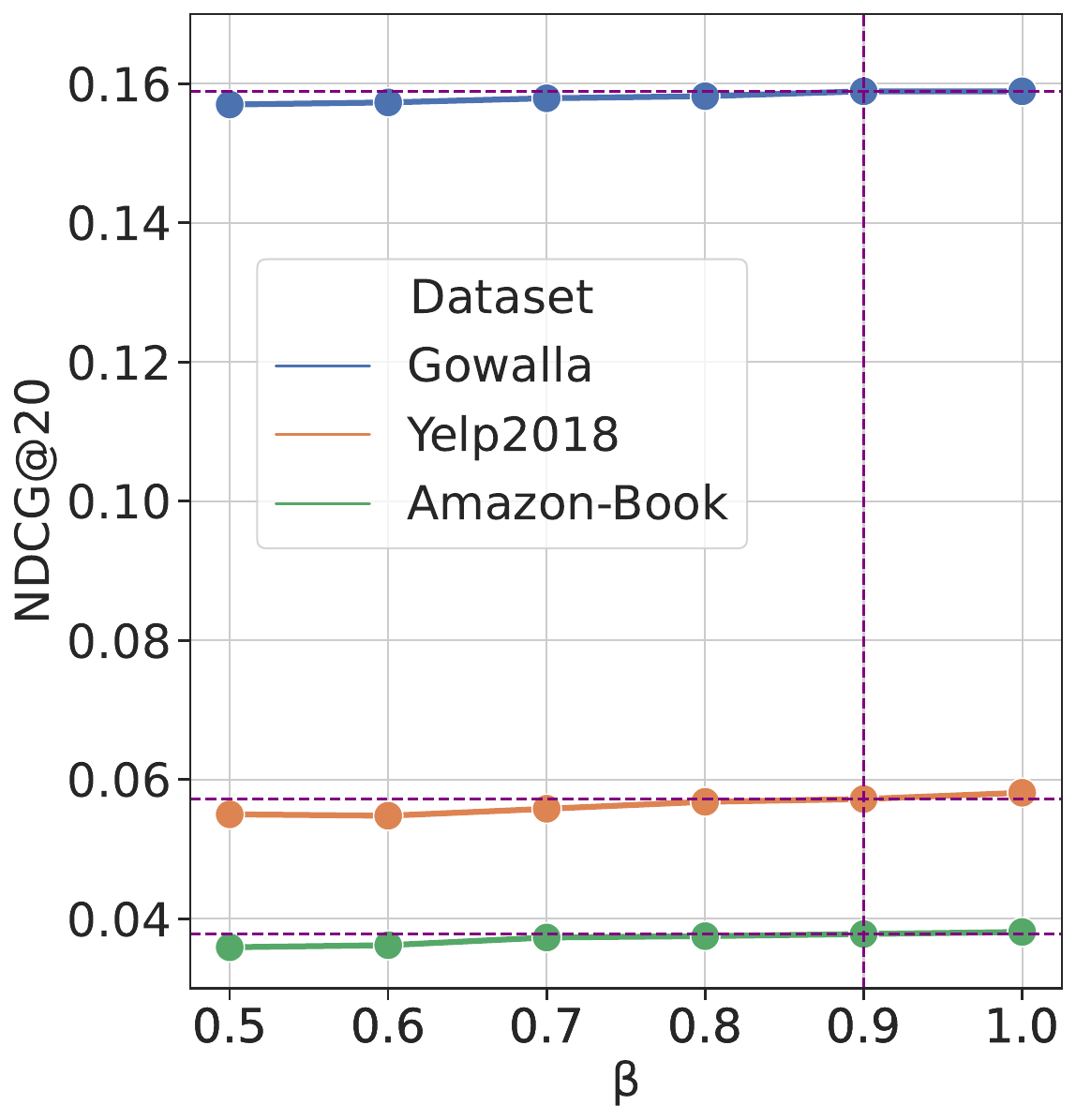}
}
\hspace{5mm}
\centering
\subfigure[Impact of $K$\label{fig:impact_param_k}]{
\includegraphics[width=1.8in]{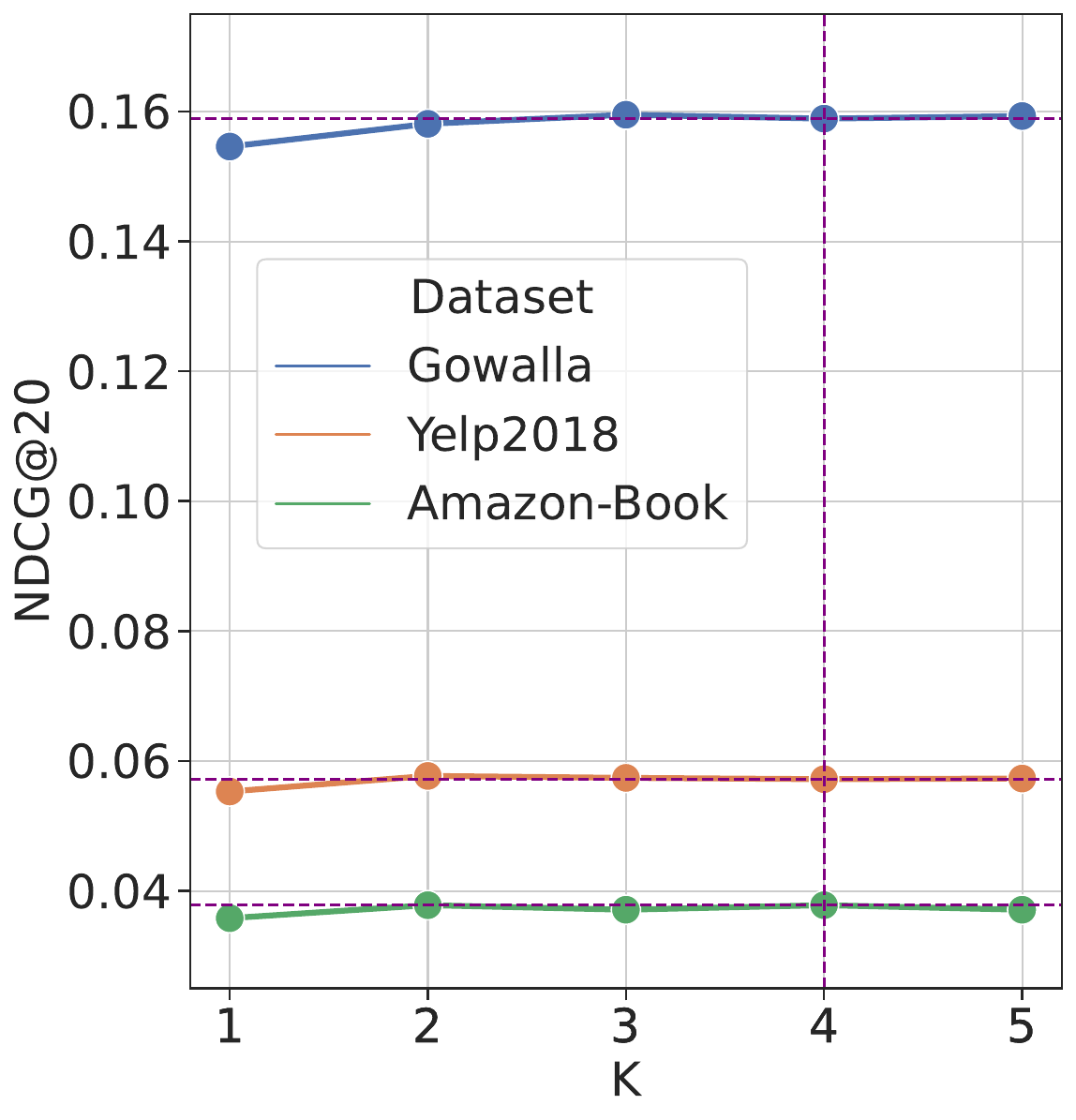}
}  
\vspace{-2mm}
\caption{
Impact of $\alpha$, $\beta$, and $K$ for Hetero-MGDCF (NDCG@20).
The vertical dashed purple lines represent the default hyperparameter setting ($\alpha=0.1$, $\beta=0.9$, and $K=4$), and the horizontal dashed purple line represents the performance under it.
\label{fig:impact_param_alpha_beta_k}
}
\end{figure*}

\subsubsection{Impact of Homogeneous Graph Construction for Homo-MGDCF}

Homo-MGDCF constructs a sparse homogeneous item-item graph via sparsification, and we use $s\%$ to denote the percentage of edges removed by sparsification.
We vary $s\%$ from $95.0\%$ to $99.9999\%$ and report the performance of Homo-MGDCF in Fig.~\ref{fig:sparsification}, where the blue lines represent the performance of Hetero-MGDCF.
The results show that Homo-MGDCF can achieve competitive performance with Hetero-MGDCF with sparse item-item graphs of $s\% \leq 97\%$.
As we remove more edges by increasing $s\%$ from $97.0\%$ to $99.9999\%$, the performance of Homo-MGDCF becomes worse than Hetero-MGDCF on the Gowalla and Yelp2018 datasets.
However, on the Amazon-Book dataset, the performance initially increases, reaching a peak when $s\%=99.9\%$, before declining as $s\%$ varies from $99.9\%$ to $99.9999\%$.
When $s\% \geq 99.999\%$, the performance of Homo-MGDCF on Amazon-Book is significantly worse than that of Hetero-MGDCF, which is similar to the scenario when $s\% \geq 98\%$ on Gowalla and Yelp2018.
The decline in performance may be attributed to the loss of meaningful edges due to the exceedingly high $s\%$.
Interestingly, Homo-MGDCF outperforms Hetero-MGDCF with significant improvement when $s\%=99.9\%$.
This shows that our sparsification approach has the potential ability to suppress noisy relational information with a proper setting of $s\%$.

\subsubsection{Impact of $\alpha$, $\beta$, and $K$}

We conduct experiments to demonstrate the impact of three primary hyperparameters ($\alpha$, $\beta$, and $K$) on Hetero-MGDCF.
Given the default parameter settings, where $\alpha=0.1$, $\beta=0.9$, and $K=4$, we vary each hyperparameter individually while keeping others fixed and report the performance to illustrate the impact of each one.
Specifically, we adjust $\alpha$ from 0.1 to 0.6 and $\beta$ from 0.5 to 1.0, around their default settings.
For $K$, we vary it from 1 to 5.
Fig.~\ref{fig:impact_param_alpha_beta_k} presents the results.
The vertical dashed purple lines represent the default hyperparameter settings ($\alpha=0.1$, $\beta=0.9$, and $K=4$), and the horizontal dashed purple line represents the performance under these settings.

In terms of $\alpha$ and $\beta$, as depicted in Fig.~\ref{fig:impact_param_alpha} and \ref{fig:impact_param_beta}, there are only negligible differences in performance when $\alpha$ or $\beta$ deviates from the default settings of $\alpha=0.1$ and $\beta=0.9$. 
Performance drops are typically observed only when $\alpha \geq 0.5$ or $\beta \leq 0.6$, values that significantly differ from the default settings. 
Regarding $K$, as shown in Fig.~\ref{fig:impact_param_k}, the performance is consistent for most values of $K$.
The sole exception is when $K=1$. 
This is consistent with CF baselines NGCF~\cite{DBLP:conf/sigir/Wang0WFC19}, LightGCN~\cite{DBLP:conf/sigir/0001DWLZ020}, and SGL-ED~\cite{DBLP:conf/sigir/WuWF0CLX21}, which also adopt multi-layer GNNs and perform poorly with only one GNN layer.
The results indicate that our model is stable and not sensitive to changes in hyperparameters.

\section{Conclusion}

In this paper, we show the equivalence between some state-of-the-art GNN-based CF models and a traditional 1-layer NRL model based on context encoding.
Based on a Markov process that trades off two types of distances, we present Markov Graph Diffusion Collaborative Filtering (MGDCF) to generalize some state-of-the-art GNN-based CF models.
Instead of considering the GNN as a trainable black box that propagates learnable user/item vertex embeddings, we treat GNNs as an untrainable Markov process that can construct constant context features of vertices for a traditional NRL model that encodes context features with a fully-connected layer.
Such simplification can help us to better understand how GNNs benefit CF models.
In particular, we show that the impact of GNNs may come from optimization rather than regularization.
This insight inspires several points for designing future GNN-based CF models. 
For example, we might explore ways to combine the optimization benefits of GNNs with explicit regularization techniques on learned user/item embeddings. 
In designing these regularization techniques, special attention should be given to the challenges posed by learnable input embeddings, which typically complicate the application of regularizations.


\vspace{-2mm}
\section*{Acknowledgments}
This research is supported by the National Research Foundation Singapore under its AI Singapore Programme (Award Number: AISG2-TC-2021-002); the National Natural Science Foundation of China (Grants 62106262, 62036012, 62072456, 62276257, 62376196, and 62206137); the SMP-IDATA Open Youth Fund; the Beijing Natural Science Foundation (Grant JQ23018); and the Tianjin Natural Science Foundation (Grant 22JCYBJC00030).


\ifCLASSOPTIONcaptionsoff
  \newpage
\fi



\bibliographystyle{IEEEtran}
\bibliography{MGDCF}
\end{document}